\documentclass[a4paper,11pt]{article}
\pdfoutput=1 % if your are submitting a pdflatex (i.e. if you have
\usepackage{jheppub} % for details on the use of the package, please
\usepackage[T1]{fontenc} % if needed
\usepackage[table]{xcolor}
\usepackage{float}

\usepackage{mathtools,amsmath,amsthm,amssymb,physics}
\usepackage{graphicx} % Required for inserting images
\usepackage{subcaption}
\usepackage{extarrows}
\usepackage{orcidlink}
\usepackage{pgffor}
\def\nn{\nonumber \\}

\makeatletter
\newlength{\apb@width}
\newcommand{\autoparbox}[2][c]{\settowidth{\apb@width}{#2}\parbox[#1]{\apb@width}{#2}}
\newcommand{\includegraphicsbox}[2][]{\autoparbox{\includegraphics[#1]{#2}}}
\makeatother

\definecolor{green1}{HTML}{244819}
\definecolor{cyan1}{HTML}{37cdaa}
\definecolor{blue1}{HTML}{5d7ac4}
\definecolor{red1}{HTML}{d0482a}
\definecolor{purple1}{HTML}{845ea8}
\definecolor{orange1}{HTML}{e07229}

\newcommand{\defas}{:=}

\newcommand{\newf}{F_1}
\newcommand{\newn}{n_1}

\newcommand{\brk}[1]{(#1)}

\newcommand{\bigbrk}[1]{\bigl(#1\bigr)}

\newcommand{\sbrk}[1]{[#1]}

\newcommand{\namedref}[2]{\hyperref[#2]{#1~\ref*{#2}}}

\makeatletter
\def\mr@ignsp#1 {\ifx\:#1\@empty\else #1\expandafter\mr@ignsp\fi}%
\newcommand{\multiref}[1]{\begingroup%\let\protect\string%
\xdef\mr@no@sparg{\expandafter\mr@ignsp#1 \: }%
\def\mr@comma{}%
\@for\mr@refs:=\mr@no@sparg\do{\mr@comma\def\mr@comma{,\,}\ref{\mr@refs}}%
\endgroup}
\makeatother
\renewcommand{\eqref}[1]{(\multiref{#1})}

\preprint{{\raggedleft%
ZU-TH 42/24
}}

\title{Reduction to master integrals and transverse integration identities}
\author[a,\,\orcidlink{0000-0001-7067-0315}]{Vsevolod Chestnov}
\author[b,\,\orcidlink{0000-0002-7543-7376}]{Gaia Fontana}
\author[a,\, \orcidlink{0000-0003-1534-4378}]{Tiziano Peraro}
\date{April 2024}

\affiliation[a]{Dipartimento di Fisica e Astronomia, Universit\'{a} di Bologna e INFN, Sezione di Bologna, via
  Irnerio 46, I-40126 Bologna, Italy}
\affiliation[b]{Physik-Institut, Universit\"{a}t Z\"{u}rich, Winterthurerstrasse 190, CH-8057 Z\"{u}rich, Switzerland}
\emailAdd{vsevolod.chestnov@unibo.it}
\emailAdd{gaia.fontana@physik.uzh.ch}
\emailAdd{tiziano.peraro@unibo.it}
\abstract{
  The reduction of Feynman integrals to a basis of linearly independent master integrals is a pivotal step in loop calculations, but also one of the main bottlenecks. In this paper, we assess the impact of using transverse integration identities for the reduction to master integrals. Given an integral family, some of its sectors correspond to diagrams with fewer external legs or to diagrams that can be factorized as products of lower-loop integrals. Using transverse integration identities, i.e.\ a tensor decomposition in the subspace that is transverse to the external momenta of the diagrams, one can map integrals belonging to such sectors and their subsectors to (products of) integrals belonging to new and simpler integral families, characterized by either fewer generalized denominators, fewer external invariants, fewer loops or combinations thereof. Integral reduction is thus drastically simpler for these new families. We describe a proof-of-concept implementation of the application of transverse integration identities in the context of integral reduction.  We include some applications to cutting-edge integral families,  showing significant improvements over traditional algorithms.

  }

\begin{document}

\maketitle
\flushbottom

\section{Introduction}
\label{sec:intro}
Feynman integrals in dimensional regularization, which are the building blocks of higher-order theoretical predictions, obey well-known linear relations, namely Integration-By-Parts (IBP) identities~\cite{Tkachov:1981wb,Chetyrkin:1981qh}, Lorentz invariance identities~\cite{Gehrmann:1999as} and symmetry relations. A crucial ingredient of multi-loop calculations is the use of such identities to reduce amplitudes and form factors to a basis of independent Feynman integrals, called \emph{master integrals} (MIs). This basis is always finite~\cite{Smirnov:2010hn} at any given loop order. However, the reduction to MIs -- also known as IBP reduction -- can become very daunting, especially at higher loops or for processes with many physical scales. In particular, reductions for cutting-edge processes at two and more loops, involving a large number of external legs or multiple scales, are a major bottleneck which often prevents important theoretical studies from being carried out. Given the key role of IBP reduction in amplitude calculations, it is essential that we develop efficient techniques to deal with its huge computational complexity.

The standard approach for the reduction to master integrals is the Laporta algorithm~\cite{Laporta:2000dsw}.  It consists in collecting identities into a large and sparse linear system, whose solution yields the list of master integrals and the reduction rules for the required set of Feynman integrals. Thanks to its implementation in private and public codes (see e.g.~\cite{vonManteuffel:2012np,Smirnov:2008iw,Maierhofer:2017gsa,Peraro:2019svx}) this approach was employed in many higher-order calculations. The solution of such a complex system, however, poses a significant computational challenge, especially because of the appearance of large intermediate expressions. Recent developments introduced the use of finite fields and rational reconstruction techniques~\cite{vonManteuffel:2014ixa,Peraro:2016wsq,Klappert:2019emp} to sidestep this problem. These approaches carry out all the operations numerically over machine-sized integers modulo a prime number and analytically reconstruct only the final result.  The latter is typically orders of magnitude less complex than the intermediate steps. These techniques are implemented in several public programs~\cite{Maierhofer:2017gsa,Peraro:2019svx,Klappert:2019emp,Smirnov:2019qkx,Klappert:2020nbg} and proved to be extremely successful in pushing the state of the art of loop calculations, producing cutting-edge results in the last years.

Despite this, IBP reduction is still one of the most challenging aspects of modern theoretical predictions and therefore a very active field of research.  Among recent developments are those that focus on building a simpler but equivalent system of identities for the reduction.  For this purpose, via the use of~\emph{syzygy} equations, one can find simpler relations between the integrals without higher powers of denominators~\cite{Gluza:2010ws,Larsen:2015ped,Bohm:2017qme,Wu:2023upw} (similar techniques are also employed in the context of multi-loop numerical unitarity~\cite{Ita:2015tya}).  Another recent development is the reconstruction of a system with a simpler block-triangular form~\cite{Guan:2024byi, Guan:2019bcx}.  These techniques often provide very substantial improvements in the efficiency of the reduction that can be critical to the feasibility of a theoretical prediction.  There has been also a recent interest in alternative approaches that completely sidestep the appearance of the large system of identities and carry out the decomposition in terms of suitably-defined scalar products and projections, such as intersection theory~\cite{Frellesvig:2019kgj, Frellesvig:2019uqt, Frellesvig:2020qot, Fontana:2023amt, Crisanti:2024onv, Fontana:2022yux, Brunello:2023rpq}. While intersection theory has many merits and opened up important connections and lines of research in mathematics and science, from the point of view of performance it is not competitive with modern implementations of the Laporta method, at the time of writing. The variety of techniques that have been proposed showcases the importance of obtaining new ways of simplifying the solution of the Laporta system, an issue that is becoming more and more crucial as the expansion in the perturbative series reaches higher orders.

In this paper, we study the impact of \emph{transverse integration} (TI) identities on the solution of the Laporta system of equations, and assess the performance improvement they can provide. Starting from an integral family contributing to a process, we identify integrals corresponding to diagrams with either fewer external legs than the original family, or that can be factorized into lower-loop integrals. These can be mapped to simpler integral families having fewer invariants or fewer loops, making the identities for these new families much simpler. We map these integrals to the new families via a transverse integration, namely a tensor decomposition in the subspace that is orthogonal to their external legs. While this technique already appeared, e.g.\ in the context of integrand decomposition~\cite{Mastrolia:2016dhn} and multi-loop numerical unitarity~\cite{Abreu:2017xsl,Abreu:2020xvt}, its impact on a traditional IBP reduction is unexplored.

In this paper, TI identities, combined with the traditional Laporta identities, are used for the reduction to master integrals. We provide a proof-of-concept implementation of the reduction to master integrals including TI identities and we assess the benefits of this approach against traditional methods on several cutting-edge examples. We stress that the use of TI identities is also compatible with more modern techniques, such as syzygy equations, but a detailed study of their combination is left to future work.

The paper is structured as follows.  In section~\ref{sec:background} we review some basic notation and definitions that are used in this work.  In section~\ref{sec:pedagogical} we present the idea of transverse integration via a pedagogical example. The method is then described in general terms in section~\ref{sec:general}, while in section~\ref{sec:applications} we discuss our proof-of-concept implementation and give benchmarks for several examples. We draw our conclusions in section~\ref{sec:conclusions}. We describe some simple algorithms for analyzing properties of integrals relevant for TI in appendix~\ref{app:algsTI} and provide more details about the TI of the integral families in our examples in appendix~\ref{app:TIfams}.

\section{Background and notation}
\label{sec:background}
In this section we set the notation and review some basic concepts and
definitions of loop integrals and IBP reduction, which will be used in the remainder of this work.

In dimensional regularization, Feynman integrals contributing at $\ell$ loops to a process with $e+1$ external momenta (only $e$ of which are linearly independent, because of momentum conservation) are integrals over the components of $\ell$ $D$-dimensional loop momenta.
It is common to organize these into integral families.  An \emph{integral family} $F$ is defined by a list of \emph{generalized denominators}
\begin{equation}
    F \leftrightarrow \{D_{F,1}, \ldots , D_{F,n}\}
    \label{eq:gen_den}
\end{equation}
that are polynomials in the loop momenta.  An integral $I_{F;\vec{a}}$ belonging to the integral family $F$ is a linear combination of integrals having the form
\begin{align}
  I_{F;\vec{a}}[N] ={}& I_{F;a_1\cdots a_n}[N] = \int \prod_{j=1}^\ell d^D k_j \ \frac{N}{\prod_{j=1}^n D_{F,j}^{a_j}}, \nn
  I_{F;\vec{a}} ={}& I_{F;a_1\cdots a_n} \equiv I_{F;\vec{a}}[1], \label{eq:int}
\end{align}
where the exponents $a_j$ are integers and $N$ can be any \emph{polynomial function} of the components of $k_1,\ldots,k_\ell$.

For any integral, we can split the set of generalized denominators into two subsets: we define \emph{proper denominators} the $D_{F,j}$ such that $a_j>0$ (i.e.\ those contributing to the denominator of the integrand) and \emph{irreducible scalar products} (\emph{ISPs}) the ones such that $a_j\leq 0$ (i.e.\ those either contributing to the numerator of the integrand or missing altogether).
For loop integrals, the generalized denominators have either the quadratic form
\begin{equation}
    D_{F,j} = l_j^2 - m_j^2,
    \label{eq:gen_den_2}
\end{equation}
or the bi-linear form
\begin{equation}
    D_{F,j} = l_j\cdot v_j - m_j^2,
    \label{eq:gen_den_1}
\end{equation}
where $l_j$ are linear combinations of loop momenta $k_1,\ldots,k_\ell$ and external momenta $p_1,\ldots,p_e$, while $m_j$ are internal masses and $v_j$ are linear combinations of external momenta.  In our applications, we focus on loop integrals contributing to scattering amplitudes, where bi-linear generalized denominators can only appear as ISPs.

Following these definitions, we can invert the relations~\eqref{eq:gen_den_2, eq:gen_den_1} to express all scalar products involving at least one
loop momentum as a linear combination of
generalized denominators $D_{F,j}$.  The requirement that this inversion is possible fixes the number $n$ of generalized denominators of a family to be
\begin{equation}
    n = \ell e + \frac{\ell (\ell+1)}{2}.
\end{equation}
Moreover, techniques such as integrand
reduction and tensor decomposition allow one to map
integrals with generic $N$ to integrals with $N=1$. Hence, when considering integrals for IBP reduction, it suffices to
focus on integrals with $N=1$.

To organize a reduction, it is convenient to partition the integrals within a family $F$ into subsets called \emph{sectors}.  A sector is
identified by the list of denominator labels $j$ such that
$a_j > 0$.  In other words, two integrals belonging to the same sector have the same list of proper denominators, possibly raised to different powers and with different numerators.  In a given sector, the~\emph{corner integral}
$I_{F;\vec{a}}$ corresponds to the only integral of that sector whose
exponents $a_j$ satisfy $a_j \in \{0,1\}, \forall j$.  In the following we will use the common convention of identifying a sector with the list exponents of its corner integral, with the notation
\begin{equation}
  S_{F;\vec{a}}=S_{F;a_1\cdots a_n},\qquad a_j\in \{0,1\},
\end{equation}
standing for the sector whose corner integral is $I_{F;\vec{a}}$.

The partition into sectors allows to establish a partial order relation to compare two sectors. A sector $S_{F;\vec{a}}$ is said to be a \emph{subsector} of a different sector $S_{F;\vec{b}}$ if its set of non-vanishing exponents $\vec{a}$ is a subset of the set $\vec{b}$ in the second sector.  In other words, $S_{F;\vec{a}}$ is a subsector of $S_{F;\vec{b}}$ if they are not the same sector and $a_j\leq b_j$ for all $j$.  In this case, we equivalently say that $S_{F;\vec{b}}$ is a \emph{parent sector} of $S_{F;\vec{a}}$.
Integrals belonging to subsectors are generally regarded as simpler than those in their parent sectors, although exceptions may exist.

In general, given an integral family $F$, only sectors with some specific combinations of proper denominators are considered, e.g.\ those that can appear in an amplitude or form factor.  For this purpose, each family is characterized by one or more \emph{top sectors}.  Thus the list of sectors of a family is restricted to its top sectors and their subsectors.

\emph{Sector mappings} are symmetries of the integrals, typically obtained via shifts of the loop momenta, which map integrals of different sectors into each other. Through these mappings, one can identify a list of \emph{unique sectors} such that any other sector of the family can be mapped into the unique ones.

To most sectors within a family (actually, all those contributing to an amplitude), we can associate a Feynman diagram, specifically the Feynman diagram of a scalar theory that is proportional to its corner integral.  We define the external legs of a sector to be those of the corresponding diagram.  In general, these will be linear combinations of the legs that define the integral family.  In the examples of this paper, we only consider families $F$ with a single top sector and we associate to them the diagram of the top sector itself.

Integrals belonging to the same family are not all linearly independent. As already stated, they obey linear relations such as integration-by-parts (IBP) identities, symmetry relations and Lorentz invariance identities. These identities allow one to reduce the initial set of integrals to a minimal, linearly independent set of master integrals.  This is also know as \emph{IBP reduction}.

The reduction to master integrals consists in systematically rewriting more complex integrals as linear combinations of simpler ones, by means of the linear identities they satisfy.  A measure of this complexity is known as \emph{weight} and defines a total ordering for a list of Feynman integrals.  While there is no unique definition of weight, the most common ones make use of the following positive integers that can be associated to a generic Feynman integral $I_{F;\vec{a}}$:
\begin{itemize}
\item the number $t$ of proper denominators
  \begin{equation}
    t \equiv \sum_{j|a_j>0} 1,
  \end{equation}
\item the total power $r$ of proper denominators
  \begin{equation}
    r \equiv \sum_{j|a_j>0} a_j,
  \end{equation}
\item the total power $s$ of ISPs
  \begin{equation}
    s \equiv -\sum_{j|a_j\leq 0} a_j,
  \end{equation}
\item the number $u$ of \emph{dots} or \emph{higher-powers of denominators}
  \begin{equation}
    u \equiv r-t.
  \end{equation}
\end{itemize}
In general, the higher these integers are, the more complex the integral $I_{F;\vec{a}}$ is regarded.  The complexity of a reduction grows, among other things, with the number of generalized denominators and the number of independent invariants which describe the process.

The reduction to master integrals is conventionally performed via the Laporta method~\cite{Laporta:2000dsw}. It consists in first writing identities for integrals $I_{F;\vec{a}}$ with symbolic exponents $a_j$, the so-called~\emph{template equations}. Then, the equations are evaluated for a given set of integer numerical values, or~\emph{seeds}, for the exponents, yielding a large and sparse linear system to be solved for the integrals. The solution yields both the list of master integrals and the reductions of the initial set of integrals.

\section{The main idea in a pedagogical example}
\label{sec:pedagogical}

\begin{figure}
    \centering
    \includegraphicsbox{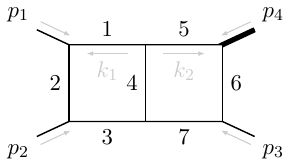}
    \caption{
        The double-box integral family with one massive leg. The labels correspond to the generalized denominators listed
        in~\protect\eqref{eq:db_props}. The thick line represents a massive, or off-shell, external leg.
    }
    \label{fig:dbox1m}
\end{figure}

In this section, we illustrate the main concepts that are explored in this
paper by means of a simple example.  We consider the two-loop double-box
integral family~\texttt{db} with one massive external leg and top sector $S_{\texttt{db};111111100}$, i.e.\ all integrals with exponents $a_{8,9}\leq0$.  It is diagrammatically shown in figure~\ref{fig:dbox1m}. This integral family has $e=3$ independent
external legs and depends on $n=9$ generalized denominators, namely 7
proper denominators and 2 ISPs for its top sector, which we define as
%
% \begin{align}
%     D_{\texttt{db}, 1} &= k_1^2
%     \\
%     D_{\texttt{db}, 2} &= \brk{k_1+p_1}^2
%     \\
%     D_{\texttt{db}, 3} &= \brk{k_1+p_1+p_2}^2
%     \\
%     D_{\texttt{db}, 4} &= \brk{k_1+k_2}^2
%     \\
%     D_{\texttt{db}, 5} &= k_2^2
%     \\
%     D_{\texttt{db}, 6} &= \brk{k_2-p_1-p_2-p_3}^2
%     \\
%     D_{\texttt{db}, 7} &= \brk{k_2-p_1-p_2}^2
%     \\
%     D_{\texttt{db}, 8} &= k_2 \cdot p_1
%     \\
%     D_{\texttt{db}, 9} &= k_1 \cdot \brk{-p_1-p_2-p_3}
% \end{align}
%
\begin{alignat}{3}
    &D_{\texttt{db}, 1} = {} k_1^2,\quad
    &
    &D_{\texttt{db}, 2} = {} \brk{k_1+p_1}^2,\quad
    &
    &D_{\texttt{db}, 3} = {} \brk{k_1+p_1+p_2}^2,
    \nonumber\\
    &D_{\texttt{db}, 4} = {} \brk{k_1+k_2}^2,\quad
    &
    &D_{\texttt{db}, 5} = {} k_2^2,\quad
    &
    &D_{\texttt{db}, 6} = {} \brk{k_2-p_1-p_2-p_3}^2,
    \nonumber
    \\
    &D_{\texttt{db}, 7} = {} \brk{k_2-p_1-p_2}^2,
    \quad
    &
    &D_{\texttt{db}, 8} = {} k_2 \cdot p_1,\quad
    &
    &D_{\texttt{db}, 9} = {} k_1 \cdot \brk{-p_1-p_2-p_3}.
    \label{eq:db_props}
\end{alignat}
Its kinematics are described by $3$ massless and one massive external momenta and we have $3$ independent
external invariants, as follows
\begin{equation}
    p_1^2 = p_2^2 = p_3^2 = 0,
    \qquad
   s = (p_1+p_2)^2,
    \qquad
    t = (p_1+p_3)^2,
    \qquad
    m^2 = p_4^2
    .
\end{equation}
The coefficients of a reduction to MIs are thus rational functions of the space-time dimension $D$ and the three invariants above.

\subsection{Integrals with fewer external legs}

\begin{figure}
    \centering
    \includegraphicsbox{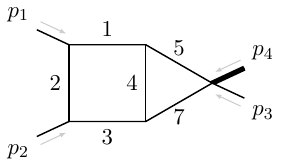}
    \quad
    $\xLongrightarrow{\hspace{1cm}}$
    \quad
    \includegraphicsbox{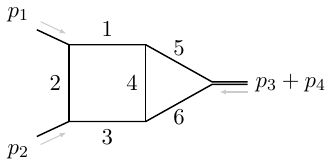}
    \caption{
        Left: sector of the double-box integral family that
        defines a new family with fewer independent external legs
        in~\protect\eqref{eq:db_sector_bt}.
        Right: the box-triangle integral family.
        Labels correspond to the generalized denominators listed
        in~\protect\eqref{eq:bt_props}.
    }
    \label{fig:bt}
\end{figure}

% \begin{figure}
%     \centering
%     \caption{
%     }
%     \label{fig:bt}
% \end{figure}

We first focus on integrals belonging to the sector $S_{\texttt{db};111110100}$ and its subsectors, i.e.\ integrals of this family with the exponents
\begin{equation}
  a_{6,8,9}\leq 0. \label{eq:alpha689leq0}
\end{equation}
The denominators that define $S_{\texttt{db};111110100}$ correspond to the
diagram displayed in figure~\ref{fig:bt}, on the left.
Integrals belonging to this sector have 6 proper denominators and 3 ISPs.
However, it is clear that the corresponding diagram only has $e=2$ independent
external legs, e.g.\ $p_1$ and $p_2$.  Indeed, the denominators of this sector
can be used to define a new integral family with only $e=2$ independent legs,
henceforth referred to as the box-triangle family \texttt{bt}
shown in figure~\ref{fig:bt}, on the right. Its generalized denominators can be chosen as
%
% \begin{align}
%     D_{\texttt{bt}, 1} &= k_1^2
%     \\
%     D_{\texttt{bt}, 2} &= \brk{k_1+p_1}^2
%     \\
%     D_{\texttt{bt}, 3} &= \brk{k_1+p_1+p_2}^2
%     \\
%     D_{\texttt{bt}, 4} &= \brk{k_1+k_2}^2
%     \\
%     D_{\texttt{bt}, 5} &= k_2^2
%     \\
%     D_{\texttt{bt}, 6} &= \brk{k_2-p_1-p_2}^2
%     \\
%     D_{\texttt{bt}, 7} &= k_2 \cdot p_2
% \end{align}
%
\begin{alignat}{3}
    &D_{\texttt{bt}, 1} = {} k_1^2,\quad
    &
    &D_{\texttt{bt}, 2} = {} \brk{k_1+p_1}^2,
    \quad
    &
    &D_{\texttt{bt}, 3} = {} \brk{k_1+p_1+p_2}^2,\nonumber\\
    &D_{\texttt{bt}, 4} = {} \brk{k_1+k_2}^2,\quad
    &
    &D_{\texttt{bt}, 5} = {} k_2^2,\quad
    &
    &D_{\texttt{bt}, 6} = {} \brk{k_2-p_1-p_2}^2,
    \nonumber\\
    &D_{\texttt{bt}, 7} = {} k_2 \cdot p_2.
   \label{eq:bt_props}
\end{alignat}
We clearly have that
\begin{equation}
    I_{\texttt{db};111110100} = I_{\texttt{bt};1111110}
    \label{eq:db_sector_bt}
\end{equation}
and more generally
\begin{equation}
  I_{\texttt{db};a_1 a_2 a_3 a_4 a_5 a_6 a_7 a_8 a_9} = I_{\texttt{bt};a_1 a_2 a_3 a_4 a_5 a_7 0}[D_{\texttt{db},6}^{-a_6} D_{\texttt{db},8}^{-a_8} D_{\texttt{db},9}^{-a_9}]  \label{eq:4to3integr}
\end{equation}
when eq.~\eqref{eq:alpha689leq0} holds. Indeed, because of eq.~\eqref{eq:alpha689leq0}, the argument in
square brackets is a polynomial numerator, hence all integrals on the
r.h.s.\ belong to sector $S_{\texttt{bt};1111110}$ or its
subsectors.

The integral family \texttt{bt} has only 7 generalized denominators (while \texttt{db} has 9) and its top sector
$S_{\texttt{bt};1111110}$ has 6 proper denominators (just as sector
$S_{\texttt{db};111110100}$) but only 1 ISP (while $S_{\texttt{db};111110100}$ has 3).  Moreover, \texttt{bt} only depends
on one external invariant, namely $s=(p_1+p_2)^2$.  Hence,
reductions for the sector $S_{\texttt{bt};1111110}$ and subsectors are
significantly simpler than those for the sector
$S_{\texttt{db};111110100}$ and subsectors, although the two are
clearly related by eq.~\eqref{eq:4to3integr}.  For the reasons mentioned in
section~\ref{sec:background}, all integrals on the r.h.s.\ of
eq.~\eqref{eq:4to3integr} can be mapped to linear combinations of integrals of
the form $I_{\texttt{bt};a_1 a_2 a_3 a_4 a_5 a_7 \beta_8}$ with integer
$\beta_8\leq0$ via a tensor decomposition.  Hence, an improvement over the
traditional approach to IBP reduction consists in mapping all integrals
belonging to the sector $S_{\texttt{db};111110100}$ and its subsectors to integrals
of the new family~\texttt{bt}, which makes their reduction drastically simpler,
for both combinatorial reasons, having fewer generalized denominators, and
because \texttt{bt} depends on fewer invariants.  A practical way to achieve
this is the method known as \emph{transverse decomposition} or \emph{transverse
integration}\footnote{Indeed, these identities can also be obtained via angular
integrations over the transverse space, see e.g.~\cite{Mastrolia:2016dhn} for
details.  In this paper, instead, we review the tensor decomposition approach.}
which is a tensor decomposition in the subspace that is transverse to the
external momenta of the new family~\cite{Mastrolia:2016dhn}.  We will now
illustrate how this works in this specific example and give a more general
description in section~\ref{sec:general}.

Any $D$-dimensional vector $v^\mu$ can be decomposed into a parallel component $v_{\parallel}^\mu$ and a transverse one $v_{\perp}^\mu$
\begin{equation}
  v^\mu = v_{\parallel}^\mu + v_{\perp}^\mu, \label{eq:vperp12}
\end{equation}
where $v_{\parallel}^\mu$ is the projection of $v^\mu$ into the subspace spanned by $p_1$ and $p_2$, that are the external legs of our new family~\texttt{bt}.  More precisely
\begin{equation}
  v_{\parallel}^\mu = c_1\, p_1^\mu + c_2\, p_2^\mu. \label{eq:vpar12}
\end{equation}
Note that
\begin{equation}
  v_{\perp}\cdot p_i = 0, \qquad v\cdot p_i = v_{\parallel}\cdot p_i\qquad \textrm{for }i=1,2. \label{eq:tdec12}
\end{equation}
By contracting eq.~\eqref{eq:vpar12} with $p_{1,2}^\mu$, one easily finds the coefficients of the decomposition
\begin{equation}
  \begin{pmatrix}
    c_1 \\ c_2
  \end{pmatrix} = \frac{2}{s}\, \begin{pmatrix}
    p_2\cdot v \\ p_1\cdot v
  \end{pmatrix}. \label{eq:cdec12}
\end{equation}
Once $v_\parallel$ is known, $v_\perp$ is simply derived from
eq.~\eqref{eq:vperp12}.  We obviously have that
\begin{equation}
  p_{1,\perp}^\mu=p_{2,\perp}^\mu=0.
\end{equation}

Let us see how to apply these identities in our example.  First, simply using
the definitions of the denominators, we can rewrite the ones appearing in the
numerator of eq.~\eqref{eq:4to3integr} as
\begin{align}
  % D_{\texttt{db}, 6} = {} & \texttt{m2 - s + Di[6] - 2 (k2 . p3) } \nn
  % D_{\texttt{db}, 8} = {} & \texttt{s/2 + Di[5]/2 - Di[6]/2 - Di[7]} \nn
  % D_{\texttt{db}, 9} = {} & \texttt{s/2 + Di[1]/2 - Di[3]/2 - (k1 . p3)}.
    D_{\texttt{db}, 6} = {} & {m^2 - s + D_{\texttt{bt}, 6} - 2 (k_2 \cdot p_3)},
    \nn
    D_{\texttt{db}, 8} = {} & {s/2 + D_{\texttt{bt}, 5}/2 - D_{\texttt{bt}, 6}/2 - D_{\texttt{bt}, 7}},
    \nn
    D_{\texttt{db}, 9} = {} & {s/2 + D_{\texttt{bt}, 1}/2 - D_{\texttt{bt}, 3}/2 - (k_1 \cdot p_3)}
    .
\end{align}
Hence, the only scalar products that are not trivially mapped to generalized denominators of family~\texttt{bt} are $(k_1\cdot p_3)$ and $(k_2\cdot p_3)$.  Therefore, we will focus on integrals of the form
\begin{equation}
  I_{\texttt{bt};\vec{a}}[(k_1\cdot p_3)^{\beta_1}\, (k_2\cdot p_3)^{\beta_2}]
\end{equation}
for integers $\beta_i\geq 0$.  As a first step, we rewrite these scalar products as
\begin{align}
  (k_1\cdot p_3) = {} & (k_1\cdot p_{3,\parallel}) + (k_{1,\perp}\cdot p_3), \nn
  (k_2\cdot p_3) = {} & (k_2\cdot p_{3,\parallel}) + (k_{2,\perp}\cdot p_3).
\end{align}
The scalar products $k_2\cdot p_{3,\parallel}$ can be easily computed using equations~\eqref{eq:tdec12} and~\eqref{eq:cdec12} which yield
\begin{equation}
  (k_i\cdot p_{3,\parallel}) = \frac{2}{s}\Big( (k_i\cdot p_1) (p_2\cdot p_3) + (k_i\cdot p_2) (p_1\cdot p_3) \Big),
\end{equation}
while $(k_i\cdot p_1)$ and $(k_i\cdot p_2)$ are rewritten as linear
combinations of the generalized denominators of the new integral
family~\texttt{bt}.  This is possible by construction, since $p_1$ and $p_2$ are external legs that define the denominators of this family, unlike $p_3$.  After this, the only scalar products that are not linear combinations of generalized denominators of~\texttt{bt} are $(k_{1,\perp}\cdot p_3)$ and $(k_{2,\perp}\cdot p_3)$.  Therefore, the r.h.s.\ of eq.~\eqref{eq:4to3integr} is a linear combination of integrals of the form
\begin{equation}
  I_{\texttt{bt};\vec{a}}[(k_{1,\perp}\cdot p_3)^{\beta_1}\, (k_{2,\perp}\cdot p_3)^{\beta_2}] = p_{3\, \mu_1}\cdots p_{3\, \mu_{\beta_1}}\, p_{3\, \nu_1}\cdots p_{3\, \nu_{\beta_2}}\, I_{\texttt{bt};\vec{a}}[k_{1,\perp}^{\mu_1\vphantom{\mu_{\beta_1}}}\cdots k_{1,\perp}^{\mu_{\beta_1}}\, k_{2,\perp}^{\nu_1\vphantom{\nu_{\beta_2}}}\cdots k_{2,\perp}^{\nu_{\beta_2}}],
\end{equation}
i.e.\ a combination of tensor integrals that only depend on the generalized denominators of the new family~\texttt{bt} and, in their numerators, on the transverse components of the loop momenta.  The tensor integrals on the r.h.s.\ can be easily expressed as combinations of integrals with no additional numerator (i.e.\ $N=1$ in eq.~\eqref{eq:int}) via a tensor decomposition
\begin{equation}
    I_{\texttt{bt};\vec{a}}[k_{1,\perp}^{\mu_1\vphantom{\mu_{\beta_1}}}\cdots k_{1,\perp}^{\mu_{\beta_1}}\, k_{2,\perp}^{\nu_1\vphantom{\nu_{\beta_2}}}\cdots k_{2,\perp}^{\nu_{\beta_2}}] = \sum_j C_j\, T_{j,\perp}^{\mu_1\cdots \mu_{\beta_1} \nu_1\cdots \nu_{\beta_2}},
    \label{eq:tenperp12}
\end{equation}
where the sum on the r.h.s.\ runs over a basis of rank-$(\beta_1+\beta_2)$
tensors that are orthogonal to $p_1$ and $p_2$, and the coefficients $C_j$ are Lorentz scalars called form factors.
This tensor decomposition is drastically simpler than a generic one, since the tensors $T_{j,\perp}$ are transverse to $p_1$ and $p_2$ and therefore can only depend on $g_\perp^{\mu\nu}$, i.e.\ the projection of the metric tensor $g^{\mu\nu}$ onto the $(D-2)$-dimensional transverse space.  This satisfies
\begin{equation}
  g_{\perp}^{\mu\nu}v_\nu = v_\perp^\mu,\qquad g_{\perp}^\mu{}_{\mu} = D_\perp = D-2.
\end{equation}
As an example we can choose for the tensor basis
\begin{equation}
    T_{1,\perp}^{\mu_1\cdots \mu_{\beta_1+\beta_2}} = g_\perp^{\mu_1 \mu_2}\, g_\perp^{\mu_3 \mu_4}\cdots  g_\perp^{\mu_{\beta_1+\beta_2-1} \mu_{\beta_1+\beta_2}},
    \label{eq:tensor_basis}
\end{equation}
and the other tensors can be obtained from $T_{1,\perp}$ via \emph{independent permutations} of its free Lorentz indexes.  The previous equations imply that the tensor integral in eq.~\eqref{eq:tenperp12} vanishes unless $\beta_1+\beta_2$ is \emph{even}.  This can also be easily proven via a change of integration variables $k_{i,\perp}^\mu\to -k_{i,\perp}^\mu$ that leaves the generalized denominators invariant and maps the tensor integral into minus itself for odd values of $\beta_1+\beta_2$.

As we will see in section~\ref{sec:general}, the form factors $C_j$ can be obtained
by contracting the tensor integral on the l.h.s.\ of~\eqref{eq:tenperp12} with
projectors $P_{j,\perp}$ that are linear combinations of the same tensors
$T_{j,\perp}$.  The coefficients of this linear combinations are rational
functions of $D_\perp$ that can be determined in a general way and only depend
on the tensor rank $\beta_1+\beta_2$ and not on the integral family (see
also~\cite{Anastasiou:2023koq,Goode:2024mci}).  By applying these projectors to the tensor
integrals, we obtain scalar products of the form $(k_{i,\perp}\cdot
k_{j,\perp})$, that in turn may be rewritten as
\begin{equation}
  (k_{i,\perp}\cdot k_{j,\perp}) = (k_i\cdot k_j) - (k_{i,\parallel}\cdot k_{j,\parallel}), \label{eq:kikjperp12}
\end{equation}
where only scalar products between loop momenta and external momenta of the new family~\texttt{bt} appear, since
\begin{equation}
  (k_{i,\parallel}\cdot k_{j,\parallel}) = \frac{2}{s}\Big( (k_i\cdot p_1) (k_j\cdot p_2) + (k_i\cdot p_2) (k_j\cdot p_1) \Big). \label{eq:kikjpar12}
\end{equation}
All of these can be written as combinations of generalized denominators $D_{\texttt{bt},j}$ of the new family~\texttt{bt}.
To make this last step more clear, we look at one explicit example.  Consider
\begin{equation}
  I_{\texttt{bt};\vec{a}}[(k_{1,\perp}\cdot p_3)\, (k_{2,\perp}\cdot p_3)] = p_{3\, \mu} p_{3\, \nu}\, I_{\texttt{bt};\vec{a}}[k_{1,\perp}^\mu\, k_{2,\perp}^\nu].
\end{equation}
The tensor decomposition reads
\begin{equation}
  I_{\texttt{bt};\vec{a}}[k_{1,\perp}^\mu\, k_{2,\perp}^\nu] = C_1\, g_\perp^{\mu\nu},
\end{equation}
where
\begin{equation}
  C_1 = \frac{1}{D-2} g_{\perp\, {\mu\nu}}\,  I_{\texttt{bt};\vec{a}}[k_{1,\perp}^\mu\, k_{2,\perp}^\nu] = \frac{1}{D-2}\, I_{\texttt{bt};\vec{a}}[(k_{1,\perp}\cdot k_{2,\perp})].
\end{equation}
Hence
\begin{equation}
  I_{\texttt{bt};\vec{a}}[(k_{1,\perp}\cdot p_3)\, (k_{2,\perp}\cdot p_3)] = \frac{1}{D-2}\, p_{3,\perp}^2 \, I_{\texttt{bt};\vec{a}}[(k_{1,\perp}\cdot k_{2,\perp})],
\end{equation}
where $p_{3,\perp}^\mu = p_{3}^\mu-p_{3,\parallel}^\mu$ and $k_{i,\perp}^\mu = k_{i}^\mu-k_{i,\parallel}^\mu$ are easily computed using the formulae above, see in particular eq.~\eqref{eq:kikjperp12} and eq.~\eqref{eq:kikjpar12}.  For the reasons explained after eq.~\eqref{eq:kikjperp12}, the argument in parentheses can be easily rewritten as a polynomial in the generalized denominators of~\texttt{bt}, yielding a linear combination of integrals of the form~$I_{\texttt{bt};\vec{a}}$ belonging to sector~$S_{\texttt{bt};1111110}$ or its subsectors.

A similar strategy can be applied to other sectors of family~\texttt{db} that correspond to diagrams with three or fewer external legs.  A more general description of the method is given in section~\ref{sec:general}.
% \begin{figure}
%     \centering
%       \begin{subfigure}{.4\textwidth}
%         \centering
%         \includegraphicsbox{figures/dbox1m2}
%         %\caption{Family \texttt{db1}. Same as \texttt{bt} in figure~\ref{fig:bt}.}
%     \end{subfigure}
%     \begin{subfigure}{.4\textwidth}
%         \centering
%         \includegraphicsbox{figures/dbox1m1}
%         %\caption{Family \texttt{db2}}
%     \end{subfigure}
%     \begin{subfigure}{.4\textwidth}
%         \centering
%         \includegraphicsbox{figures/dbox1m3}
%         %\caption{Family \texttt{db3}}
%     \end{subfigure}
%     \begin{subfigure}{.4\textwidth}
%         \centering
%         \includegraphicsbox{figures/dbox1m4}
%         %\caption{Family \texttt{db4}}
%     \end{subfigure}
%     \caption{
%         New families for transverse integration of the double box with one external mass.  All unique sectors of the double box family in figure~\ref{fig:dbox1m} with fewer than 4 external legs can be mapped, via transverse integration identities, to one of these simpler families.
%     }
%     \label{fig:dbox1m_oth_new_families}
% \end{figure}

\subsection{Factorizable integrals}
\label{sec:factorizableexample}

\begin{figure}
    \centering
    \includegraphicsbox{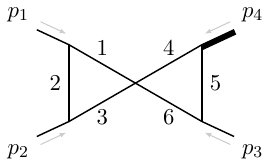}
    \caption{A factorizable sector of the double-box family.}
    \label{fig:butterfly}
\end{figure}

We now analyze the integrals belonging to the sector $S_{\texttt{db};111011100}$ and its subsectors, i.e.\ integrals of this family with the exponents
\begin{equation}
  a_{4,8,9}\leq 0. \label{eq:alpha489leq0}
\end{equation}
The denominators of this sector correspond to the diagram
shown in figure~\ref{fig:butterfly},
which is clearly the product of two one-loop diagrams.  We manifestly have that
\begin{equation}
  I_{\texttt{db};a_1 a_2 a_3 0 a_5 a_6 a_7  00} = I_{\texttt{t1};a_1 a_2 a_3}\, I_{\texttt{t2};a_5 a_6 a_7}
\end{equation}
where \texttt{t1} and \texttt{t2} are the one-loop integral families
\begin{align}
  I_{\texttt{t1};a_1 a_2 a_3} &= \int d^D k_1\, \frac{1}{D_{\texttt{t1},1}^{a_1}\, D_{\texttt{t1},2}^{a_2}\, D_{\texttt{t1},3}^{a_3}},\label{eq:factfamt1}
\\
  I_{\texttt{t2};a_5 a_6 a_7} &= \int d^D k_2\, \frac{1}{D_{\texttt{t2},5}^{a_5}\, D_{\texttt{t2},6}^{a_6}\, D_{\texttt{t2},7}^{a_7}}, \label{eq:factfamt2}
\end{align}
respectively defined by the following sets of generalized denominators
\begin{alignat}{3}
    D_{\texttt{t1}, 1} = {}& k_1^2
     \quad
    &
    D_{\texttt{t1}, 2} = {}& \brk{k_1+p_1}^2
    \quad
    &
    D_{\texttt{t1}, 3} = {}& \brk{k_1+p_1+p_2}^2,
    \nonumber\\
    D_{\texttt{t2}, 5} = {}& k_2^2,
    \quad
    &
    D_{\texttt{t2}, 6} = {}& \brk{k_2-p_1-p_2-p_3}^2,
    \quad
    &
    D_{\texttt{t2}, 7} = {}& \brk{k_2-p_1-p_2}^2.
\end{alignat}
By carrying out a tensor decomposition one loop at a time, any integral of
the form $I_{\texttt{db};\vec{a}}$ with exponents satisfying
eq.~\eqref{eq:alpha489leq0} can be rewritten as a linear combination of
integrals of the form $I_{\texttt{t1};\vec{b}_1}\, I_{\texttt{t2};\vec{b}_2}$
i.e.\ as a combination of products of lower-loop integrals.  Each of these one-loop integrals only has 3 generalized denominators and depends on fewer scales (1 and 2 respectively) with respect to the corresponding integrals in the~\texttt{db} family.  Their integral reduction is thus orders of magnitude easier than the one for sector~$S_{\texttt{db};111011100}$ of the original family~\texttt{db}.  Once again, an effective way of carrying out this tensor decomposition is the method of transverse integration.

A possible approach consists in carrying out the transverse integration one loop factor at a time.  For this purpose, we rewrite any integral satisfying~\eqref{eq:alpha489leq0} as
\begin{equation}
  I_{\texttt{db};a_1 a_2 a_3 a_4 a_5 a_6 a_7 a_8 a_9} = \int d^D k_2\, \frac{D_{\texttt{db},8}^{-a_8}}{D_{\texttt{db},5}^{a_5}\, D_{\texttt{db},6}^{a_6}\, D_{\texttt{db},7}^{a_7}} I_{\texttt{t1};a_1 a_2 a_3}[D_{\texttt{db},4}^{-a_4}\, D_{\texttt{db},9}^{-a_9}], \label{eq:k1factintegr}
\end{equation}
where, from the point of view of the innermost integral of family~\texttt{t1}, the momentum $k_2$ is just an arbitrary, off-shell reference momentum, that only appears in the numerator of the integrand, because of eq.~\eqref{eq:alpha489leq0}.

We now follow the exact same steps as in the previous subsection, while working on the innermost integration for the new family~\texttt{t1}.  First, we note that we can rewrite
\begin{align}
    % D_{\texttt{db},4} = {} & \texttt{Di[1] + 2 (k1 . k2) + (k2 . k2) } \nn
    % D_{\texttt{db},4} = {} & \texttt{s/2 + Di[1]/2 - Di[3]/2 - (k1 . p3) }
    D_{\texttt{db},4} = {} & {D_{\texttt{t1}, 1} + 2 (k_1 \cdot k_2) + (k_2 \cdot k_2) }, \nn
    D_{\texttt{db},9} = {} & {s/2 + D_{\texttt{t1}, 1}/2 - D_{\texttt{t1}, 3}/2 - (k_1 \cdot p_3) },
\end{align}
which means, remembering that anything that does not depend on $k_1$ can be pulled outside the innermost integration, we can focus on integrals of the form
\begin{equation}
  I_{\texttt{t1};a_1 a_2 a_3}[(k_1\cdot k_2)^{\beta_1}\, (k_1\cdot p_3)^{\beta_2}],
\end{equation}
with integers $\beta_i\geq 0$.  Then we proceed the exact same way as in the previous subsection, by performing a transverse integration with respect to the subspace spanned by $p_1$ and $p_2$, that are the external legs of the one-loop triangle family~\texttt{t1}.  At the end of this procedure, all the integrals in the last equation are rewritten as linear combinations of integrals $I_{\texttt{t1};\vec{a}}$.  A careful review of all steps shows that, in the coefficients of such linear combination, the momentum $k_2$ (as well as $p_3$) only appear in scalar products in the numerator.  Hence, integrals in eq.~\eqref{eq:k1factintegr} become linear combinations of integrals of the form
\begin{equation}
  I_{\texttt{t2};\vec{b}}[N(k_2)]\times I_{\texttt{t1};\vec{a}},
\end{equation}
where now the integration is fully factorized into a product of lower loop integrals.  The polynomial numerator $N(k_2)$ now contains scalar products of the form $k_2^2$ and $k_2\cdot p_i$ for $i=1,2,3$.  The denominators of \texttt{t2} however correspond to diagrams with, again, only $e=2$ independent external momenta, which in this case are $p_3$ and $p_4$.  A transverse integration with respect to the subspace spanned by $p_3$ and $p_4$ will, finally, yield a combination of integrals of the previous form but with $N(k_2)=1$, i.e.
\begin{equation}
  I_{\texttt{t2};\vec{b}}\times I_{\texttt{t1};\vec{a}}. \label{eq:factend12}
\end{equation}
At this stage, therefore, any integral belonging to sector $S_{\texttt{db};111011100}$ and its subsectors is mapped to a linear combination of integrals of the form shown in eq.~\eqref{eq:factend12}.  The IBP reduction of a lower loop integral is, in turn, orders of magnitude simpler in algebraic complexity.

In the next section, we will describe how to apply these ideas to a more general $\ell$-loop integral, while in section~\ref{sec:applications} we will show some practical applications to IBP reduction.

\section{General algorithm for transverse integration}
\label{sec:general}

In this section, we illustrate the transverse integration (TI) algorithm for a generic $\ell$-loop integral family, generalizing the discussion of section~\ref{sec:pedagogical} which instead focused on simple specific examples.

Given an integral family $F$ to be reduced, the key idea is to find sectors
whose corner integrals correspond to either diagrams with fewer external legs
than those defining $F$, or diagrams that can be factorized as products of
lower-loop integrals.  Then, using transverse integration identities, we map
integrals belonging to these sectors and their subsectors to (products of)
integrals belonging to new and simpler integral families $F_j$ characterized by
either fewer generalized denominators, fewer external invariants, lower loops or
combinations thereof.  Integral reduction is thus drastically simpler for the
newer families $F_j$. In the following we call such sectors \emph{TI-mapped sectors} and the corresponding new families \emph{TI families}.

At the core of the method is the decomposition of the $D$-dimensional Minkowski
space into a parallel and a transverse component. Consider a set of
$D_\parallel$ independent momenta
\begin{equation}\label{eq:qparallel}
    q_1^\mu,\ldots,q_{D_\parallel}^\mu,
\end{equation}
that are linear combinations of some of the external momenta $p_i^\mu$ that define a
family $F$.  We define the decomposition of a generic $D$-dimensional vector
$v^\mu$
\begin{equation}
  v^\mu = v_{\parallel}^\mu + v_{\perp}^\mu \label{eq:vperp}
\end{equation}
into a parallel component $v_{\parallel}^\mu$ that is the projection of $v^\mu$ into the subspace spanned by the momenta $q_j$
\begin{equation}
  v_{\parallel}^\mu = \sum_{j=1}^{D_\parallel} c_j\, q_j^\mu \label{eq:vpar}
\end{equation}
and a transverse component $v_\perp^\mu$ that is orthogonal to this subspace
\begin{equation}
  v_{\perp}\cdot q_i = v_\perp \cdot v_\parallel = 0 \qquad \textrm{for }i=1,\cdots,D_\parallel. \label{eq:tdec}
\end{equation}
The coefficients of the decomposition in eq.~\eqref{eq:vpar} can be written in terms of scalar products $q_j\cdot v$ as
\begin{equation}
    c_i = \sum_{j=1}^{D_\parallel} Q^{-1}_{ij}\, (q_j\cdot v)
\end{equation}
where $Q^{-1}$ is the inverse of the $D_\parallel \times D_\parallel$ Gram matrix of entries
\begin{equation}
    Q_{ij} \equiv q_i\cdot q_j.
    \label{eq:gram_def}
\end{equation}
Once $v_\parallel^\mu$ has been computed this way, $v_\perp^\mu$ can be obtained using \eqref{eq:vperp}.  In this context, we also decompose the $D$-dimensional metric tensor $g^{\mu\nu}$ into a parallel and transverse component
\begin{equation}
  g^{\mu\nu} = g_\parallel^{\mu\nu} + g_\perp^{\mu\nu}
\end{equation}
with
\begin{equation}
  g_{\perp}^{\mu\nu}v_\nu = v_\perp^\mu,\qquad g_{\perp}^\mu{}_{\mu} = D_\perp \equiv D-D_\parallel,
\end{equation}
where in the last equation we defined the transverse space-time dimension $D_\perp$.  Note that, while $D_\parallel$ is an integer, $D_\perp$ retains the full dependence on the dimensional regulator.  We also obviously have that
\begin{equation}
  q_{j,\parallel}^\mu = q_{j}^\mu, \qquad q_{j,\perp}^\mu = 0 \qquad \textrm{for }j=1,\ldots,D_\parallel.
\end{equation}
By combining the previous equations, we find
\begin{equation}\label{eq:vwparperp}
  (v\cdot w) =  (v_\parallel \cdot w_\parallel) +   (v_\perp \cdot w_\perp)
\end{equation}
with
\begin{equation}\label{eq:vwpar}
    \brk{v_\parallel, w_\parallel}
    = \sum_{ij} \brk{v, q_i} \> \bigbrk{Q^{-1}}_{ij} \> \brk{q_j, w} ,
\end{equation}
for any two vectors $v^\mu$ and $w^\mu$.  It is also worth noting that this decomposition is not possible when $D_\parallel=1$ and $q_1^2=0$, since the $1\times 1$ Gram matrix is obviously vanishing in that case.

In the following, we apply these concepts to the transverse integration of loop
integrals in the context of IBP reduction.  As in the previous section, we first focus on sectors identified by diagrams with fewer legs (subsection~\ref{sec:fewerlegs}) and then on sectors whose diagrams factorize as products of lower loops (subsection~\ref{sec:factorizable}).

\subsection{Sectors with fewer external legs}
\label{sec:fewerlegs}
Consider a generic sector $S_{F}$, with $t$ proper denominators, that corresponds to a diagram with fewer external legs than the top sector of $F$.  This defines a list of $D_\parallel$ independent external momenta $q_j$, as in eq.~\eqref{eq:qparallel}, for a non-negative integer $D_\parallel<e$.  We note that, in order to make this dependence manifest at the integrand level, a \emph{shift} in the loop momenta might be required
\begin{equation}\label{eq:kshift}
  k_i^\mu \to k_{s,i}^\mu = \sum_j\, \alpha_{ij} k_j^\mu + \sum_{ij}\beta_{ij} p_j^\mu, \qquad \alpha_{ij},\beta_{ij}\in \{0,\pm 1\}.
\end{equation}
As a simple example, consider the one-loop integral
\begin{equation}
  I=\int d^D k \, \frac{N(k)}{(k+p_1)^2\, (k+p_1+p_2)^2\, (k+p_1+p_2+p_3+p_4)^2},
\end{equation}
whose proper denominators define a diagram with $D_\parallel=2$ independent legs, e.g.\
\begin{equation}
  q_1^\mu = p_2^\mu, \qquad q_2^\mu= p_3^\mu + p_4^\mu
\end{equation}
but this is only manifest after the shift
\begin{equation}
  k^\mu \mapsto k_s^\mu = k^\mu - p_1^\mu
\end{equation}
which yields
\begin{equation}
  I=\int d^D k_s \, \frac{N(k_s)}{(k_s)^2\, (k_s+q_1)^2\, (k_s+q_1+q_2)^2}.
\end{equation}
We refer to such sectors as \emph{sectors with fewer legs}.  In general,
identifying these and the corresponding shift of loop momenta is a relatively
simple task.  We describe a purely algebraic method in appendix~\ref{sec:ffextlegs}.

Up to a relabelling of the generalized denominators, we can assume $S_{F}$ to be of the form $S_{F;1\cdots 1 0 \cdots  0}$.  Hence, integrals belonging to either this sector or its subsectors satisfy
\begin{equation}\label{eq:fewajless0}
    a_j\leq 0, \qquad j=t+1,\ldots,n.
\end{equation}
We thus define a new integral family $\newf$ with $D_\parallel$ independent external momenta and $\newn$ generalized denominators $D_{\newf,j}$, where
\begin{equation}
  \newn = \ell\, D_\parallel + \frac{\ell (\ell+1)}{2} < n,
\end{equation}
while its proper denominators coincide with the ones of sector $S_{F}$, up to the aforementioned shift
\begin{equation}
  D_{\newf;j} = D_{F;j} \Big|_{k\to k_s}, \qquad j=1,\ldots,t.
\end{equation}
As usual, the remaining generalized denominators of $\newf$ are defined such that every scalar product of the form $k_i\cdot k_j$ and $k_i\cdot q_j$ is a linear combination of the $\newn$ generalized denominators.  The top sector of this new family is
\begin{equation}
    S_{\newf} \equiv S_{
        \newf;
        \underbrace{\scriptstyle{1\cdots1}}_{t}
        \underbrace{\scriptstyle{0\cdots 0}}_{\newn-t}
        % \underbrace{{1\cdots1}}_{t\textrm{times}}
        % \underbrace{{0\cdots 0}}_{\newn-t \textrm{ times}}
    }.
\end{equation}

The $n$ generalized denominators of family $F$, after the shift in eq.~\eqref{eq:kshift}, can be rewritten as linear combinations of the $\newn$ generalized denominators of $\newf$ and $n-\newn$ additional scalar products.  More explicitly
\begin{equation}\label{eq:denFtodennewF}
  D_{F,i} \Big|_{k\to k_s} = d_{i0} + \sum_{m=1}^{\newn}d_{im}\, D_{\newf,m} + \sum_{m=1}^{n-\newn} \tilde d_{im} \tilde D_{\newf,m}
  ,
\end{equation}
with $d_{i0}$, $d_{im}$, and $\tilde d_{im}$ independent of the loop momenta.
Note that all these additional scalar products have the form
\begin{equation}\label{eq:extraisps}
  \tilde D_{\newf,m} = k_{i_m}\cdot p_{j_m},
\end{equation}
for suitable indexes $i_m$ and $j_m$, since by construction all those of the form $k_i\cdot k_j$ are already linear combinations of generalized denominators of the new family $\newf$.  Therefore, by means of straightforward manipulations on the integrands, we rewrite any integral $I_{F;\vec{a}}$ of family $F$ satisfying eq.~\eqref{eq:fewajless0} as a linear combination of integrals of the form
\begin{equation}
  I_{\newf;a_1\cdots a_{\newn}}\Big[\prod_{m=1}^{n-\newn} (k_{i_m}\cdot p_{j_m})^{\beta_{m}}\Big]
  ,
\end{equation}
for integers $\beta_m\geq 0$.  The dependence of the integrals on the additional scalar products can be removed via a tensor decomposition that is efficiently implemented using transverse integration identities.  We first rewrite every scalar product of the form $(k\cdot p)$, where $k$ can be any of the loop momenta and $p$ any of the momenta of $F$ appearing in the additional scalar products $\tilde D_{\newf,m}$ in eq.~\eqref{eq:extraisps}, using
equations~\eqref{eq:vwparperp} and~\eqref{eq:vwpar}, namely
\begin{equation}
  (k \cdot p) = \sum_{ij} (k \cdot q_i) (Q^{-1})_{ij} (q_j\cdot p) + (k_\perp \cdot p),
\end{equation}
where the Gram matrix $Q$ was defined in eq.~\eqref{eq:gram_def}.  The scalar products $(k_i\cdot q_j)$ are, once again by construction, linear combinations of generalized denominators $D_{\newf,m}$ of the new family.  By combining this identities with eq.~\eqref{eq:denFtodennewF} we thus rewrite the generalized denominators of $F$ as
\begin{equation}\label{eq:denFtodennewFort}
  D_{F,i} \Big|_{k\to k_s} = d_{i0,\perp} + \sum_{m=1}^{\newn} d_{im,\perp}\, D_{\newf,m} + \sum_{m=1}^{n-\newn} \tilde d_{im,\perp} \tilde D_{\newf,m,\perp},
\end{equation}
with $d_{i0,\perp}$, $d_{im,\perp}$, and $\tilde d_{im,\perp}$ independent of the loop momenta, while the orthogonal scalar products $\tilde D_{\newf,m,\perp}$ have the form
\begin{equation}\label{eq:extraispsort}
  \tilde D_{\newf,m,\perp} = k_{i_m,\perp}\cdot p_{j_m}
\end{equation}
and only depend on the orthogonal components $k_{j,\perp}^\mu$ of the loop momenta.  Inserting this decomposition into a generic integral of $F$ satisfying eq.~\eqref{eq:fewajless0},
yields a linear combination of integrals the form
\begin{equation}\label{eq:Ikiortpj}
  I_{\newf;a_1\cdots a_{\newn}}\Big[\prod_{m=1}^{n-\newn} (k_{\perp,i_m}\cdot p_{j_m})^{\beta_{m}}\Big]
  ,
\end{equation}
with $\beta_i\geq 0$, i.e.\ involving only generalized denominators of $\newf$ and scalar products between other external momenta and the transverse components of the loop momenta.

Similarly as before, the transverse scalar products are removed via a tensor decomposition in the transverse space
\begin{align}
    I_{\newf; \vec{a}}\sbrk{k_{i_1,\perp}^{\mu_1}\cdots k_{i_r,\perp}^{\mu_r}}
    =
    \sum_j C_j \> T_{j,\perp}^{\mu_1 \cdots \mu_r}\label{eq:tidec},
\end{align}
where $C_j$ are scalar form factors and $T_{\perp,j}^{\mu_1 \cdots \mu_r}$ is a basis of rank-$r$ covariant tensors in the $D_\perp$-dimensional transverse space. Such a basis can be chosen using
\begin{equation}
  T_{1,\perp}^{\mu_1 \cdots \mu_r} = g_{\perp}^{\mu_1\mu_2}\cdots g_{\perp}^{\mu_{r-1}\mu_{r}},
\end{equation}
while the others can be obtained from $T_{1,\perp}$ via independent permutations of the Lorentz indices (there are $r!/(2^{r / 2} (r / 2)!)$ independent tensors, without accounting for symmetries of the integrand).  Here $g_{\perp}^{\mu\nu}$ is the projection of the metric tensor into the subspace that is transverse to $q_1,\ldots,q_{D_\parallel}$, as defined above.  The form factors $C_j$ are computed by contracting the l.h.s.\ of eq.~\eqref{eq:tidec} with suitable projectors $P_{j,\perp}$ such that
\begin{equation}
      C_i =
     P_{i, \perp}^{\mu_1 \cdots \mu_r}
     \>
     I_{\newf; \vec{a}}\sbrk{k_{i_1,\perp}{}_{\mu_1}\cdots k_{i_r,\perp}{}_{\mu_r}}.
\end{equation}
The projectors $P_j$ are also linear combinations of the same tensors $T_{j,\perp}$, namely
\begin{align}
    P_{i, \perp}^{\mu_1 \cdots \mu_r}
    =
    \sum_j (T^{-1})_{ij} \> T_{j,\perp}^{\mu_1 \cdots \mu_r}
\end{align}
where $T$ is the matrix of entries
\begin{equation}
  T_{ij} \equiv T_{i,\perp}^{\mu_1 \cdots \mu_r} T_{j,\perp}\, {}_{\mu_1 \cdots \mu_r}.
\end{equation}
From these equations it should be clear that a basis of transverse tensors and the corresponding set of projectors can be computed, for each value of the rank $r$, once and for all, independently of the integral family.  In particular the matrix $T_{ik}$ and its inverse (that appears in the projectors) are univariate rational functions of $D_\perp$. Moreover, as already argued in section~\ref{sec:pedagogical}, we have
\begin{equation}
  I_{\newf; \vec{a}}\sbrk{k_{i_1,\perp}^{\mu_1}\cdots k_{i_r,\perp}^{\mu_r}}=0, \qquad \textrm{ for $r = $ odd}.
\end{equation}
For most phenomenologically relevant applications, the decomposition
in eq.~\eqref{eq:tidec} is thus easy to obtain even via a naive
inversion of the matrix $T$.  In cases where integrals with very high-rank 
numerators are needed, improved tensor reduction
techniques exist (see e.g.~\cite{Anastasiou:2023koq,Goode:2024mci})
and can effectively mitigate the complexity of this step.

By applying these projectors to our integrals we obtain a combination of integrals
of the form
\begin{equation}\label{eq:Ikiortkj}
  I_{\newf; \vec{a}}\Big[\prod_{i\leq j} (k_{i,\perp}\cdot k_{j,\perp})^{\beta_{ij}}\Big],
\end{equation}
where, using again equations~\eqref{eq:vwparperp} and~\eqref{eq:vwpar},
\begin{equation}\label{eq:kiortkjorttopar}
    (k_{i,\perp}\cdot k_{j,\perp})
    =
    (k_i \cdot k_j) - \sum_{lm} (k_i \cdot q_l) (Q^{-1})_{lm} (k_j \cdot q_m).
\end{equation}
All scalar products on the r.h.s.\ of the last equation are linear combinations of the denominators $D_{\newf,j}$ of the new family $\newf$.
The procedure we outlined thus successfully rewrites any integral belonging sector $S_{F}$ of the original family $F$ or its subsectors into a linear combination of integrals of a new family $\newf$, characterized by fewer external legs, fewer ISPs and invariants.

\subsection{Factorizable sectors}
\label{sec:factorizable}
A second kind of TI involves so-called factorizable sectors, i.e.\ sectors whose corner integral can be written as a product of lower loop integrals.  In general, in order to achieve this, a shift of the loop momenta is required.  The latter has the form in eq.~\eqref{eq:kshift} with all $\beta_{ij}=0$.  A simple and purely algebraic method of identifying such sectors and the corresponding shift is described in appendix~\ref{sec:ffloopfactors}.  All integrals belonging to the sector and its subsectors can thus be written as linear combinations of products of lower loop integrals.

Consider a factorizable sector~$S_{F;\vec{a}}$ of family $F$ and assume the aforementioned shift in the loop momenta has already been performed.  The proper denominators of such sector can be partitioned in at least two sets, according to the loop momenta they depend on.  Up to a relabeling of the denominators and the loop momenta, consider a first set of proper denominators
\begin{equation}\label{eq:denfact1}
  D_{F;1},\ldots D_{F;t_1}
  ,
\end{equation}
which only depend on the loop momenta
\begin{equation}
  k_1,\ldots k_{\ell_1}
  ,
\end{equation}
and a second set of proper denominators
\begin{equation}\label{eq:denfact2}
  D_{F;t_1+1},\ldots D_{F;t_1+t_2}
  ,
\end{equation}
which only depend on the loop momenta
\begin{equation}
  k_{\ell_1+1},\ldots k_{\ell_1+\ell_2},
\end{equation}
where $t=t_1+t_2$ and $\ell=\ell_1+\ell_2$ are the total number of
proper denominators and the loop order of $S_{F;\vec{a}}$
respectively.  We thus have that
\begin{equation}
  I_{F;a_1,\cdots a_t 0\cdots 0} = \left( \int  \prod_{j=1}^{\ell_1} d^D k_j \frac{1}{D_1^{a_1}\cdots D_{t_1}^{a_{t_1}}} \right) \left( \int \prod_{j=\ell_1+1}^{\ell_1+\ell_2} d^D k_j \frac{1}{D_{t_1+1}^{a_{t_1+1}}\cdots D_{t_1+t_2}^{a_{t_1+t_2}}} \right)
  ,
\end{equation}
for any exponent $a_j$ with $j\leq t$.  We thus define two lower-loop TI families, $F_1$ and $F_2$, whose top sectors have the proper denominators in eq.~\eqref{eq:denfact1} and eq.~\eqref{eq:denfact2} respectively.  In general, as in the previous subsection, the definition of the new TI families may require adding ISPs.  The previous equation thus reads
\begin{equation}
  I_{F;a_1,\cdots a_t 0\cdots 0} = I_{F_1;a_1,\cdots a_{t_1} 0\cdots 0} \times I_{F_2;a_{t_1+1},\cdots a_{t_1+t_2} 0\cdots 0}.
\end{equation}
Hence, the mapping is straightforward except when the ISPs of $S_{F,\vec{a}}$ have non-trivial exponents.  We observe, however, that we can write a generic integral belonging to $S_{F,\vec{a}}$ and subsectors as
\begin{equation}
  I_{F;a_1,\cdots a_t 0\cdots 0}[N(k_1,\ldots,k_\ell)] = \left(\int \prod_{j=\ell_1+1}^{\ell_1+\ell_2} d^D k_j \frac{1}{D_{t_1+1}^{a_{t_1+1}}\cdots D_{t_1+t_2}^{a_{t_1+t_2}}} \right) I_{F_1;a_1\cdots a_{t_1} 0\cdots 0}[N(k_1,\ldots,k_\ell)],
\end{equation}
where the numerator $N$ is polynomial in the ISPs of $F$ and therefore in the scalar products one can build from the loop and external momenta of $F$.  We thus focus on the $\ell_1$-loop integral $I_{F_1;a_1\cdots a_{t_1} 0\cdots 0}[N(k_1,\ldots,k_\ell)]$ and handle its numerator by following the same TI steps as in section~\ref{sec:fewerlegs}.  More specifically, $N(k_1,\ldots,k_\ell)$ may contain scalar products which cannot be written as a linear combination of generalized denominators of $F_1$.  The corresponding integrals, however, will be cast -- by means of the algorithm in~\ref{sec:fewerlegs} -- as a linear combination of integrals $I_{F;\vec{b}}$ in the family $F_1$.  Moreover, by reviewing these steps of the TI algorithm, it is easy to see that the coefficients $N_{\vec{b}}$ of such linear combination will contain, in their denominators, only scalar products between the external legs $q_1,\ldots,q_{D_\parallel}$ of sector $S_{F_1;a_1\cdots a_{t_1} 0\cdots 0}$.  Hence, the dependence of these coefficients on any other momentum which may appear in $N(k_1,\ldots,k_\ell)$, including the other loop momenta $k_{\ell_1+1},\ldots k_{\ell_1+\ell_2}$, will be in the numerator of these coefficients.  We thus obtain a relation of the form
\begin{equation}
  I_{F;a_1,\cdots a_t 0\cdots 0}[N(k_1,\ldots,k_\ell)] = \sum_{\vec{b}} I_{F_2;a_{t_1+1},\cdots a_{t_1+t_2} 0\cdots 0}[N_{\vec{b}}(k_{\ell_1+1},\ldots k_{\ell_1+\ell_2})]\times I_{F_1;\vec{b}}.
\end{equation}
Now we proceed by treating the numerators of the integrals of family~$F_2$ appearing in the previous equation using the same method we just used for $F_1$, namely by following the TI algorithm of section~\ref{sec:fewerlegs}.  The final outcome has the form
\begin{equation}
  I_{F;a_1,\cdots a_t 0\cdots 0}[N(k_1,\ldots,k_\ell)] = \sum_{\vec{b_1},\vec{b_2}}\, c_{\vec{b_1}\vec{b_2}}\, \, I_{F_2;\vec{b_2}}\times I_{F_1;\vec{b_1}},
\end{equation}
where all numerators on the r.h.s.\ are now combinations of the generalized denominators of the TI families $F_1$ and $F_2$.

The same method can be used, recursively, for factorizable sectors whose corner integral is the product of more than two sub-loop factors.

\subsection{Mappings between new families}
\label{sec:mappings}

Using the algorithms of the previous subsection, we can map many integrals of a family $F$ to combinations of integrals of simpler families $F_i$, which we call TI families, having fewer legs or fewer loops.  There often are, however, mappings between sectors belonging to different TI families $F_i$.  Taking these into account is essential for a reduction to a minimal basis of MIs.

A pragmatic way of finding these identities is by looking at how integrals of a TI family $F_i$ can be mapped back to family $F$.  Consider an integral $I_{F_i}\in F_i$.  By undoing the shift in the loop momenta -- i.e.\ by inverting eq.~\eqref{eq:kshift} -- and rewriting all scalar products in terms of the generalized denominators of $F$, we can easily map back $I_{F_i}$ to an integral $I_F$ of $F$ (or a linear combination of integrals).  By applying sector mappings of $F$ to $I_F$, we can assume that $I_F$ belongs to a unique sector of $F$.  By applying the same algorithm to many integrals $I_{F_i}$ belonging to different families $F_i$, we build a system of identities which maps them back to $F$.  This, combined with the TI identities we already found, allows us to relate integrals of different TI families by reusing sector mappings and TI identities of $F$ only.  In practice, we apply this only to the master integrals of each TI family $F_i$ belonging to sectors with fewer external legs than the top sector of $F_i$.  This yields mappings between the master integrals of different families $F_i$, allowing us to obtain, after TI, a reduction to a minimal number of master integrals.

\section{Applications}
\label{sec:applications}

In this section we apply transverse integration (TI) identities to the
reduction of highly non-trivial integral families.  We describe a relatively
simple way of incorporating these identities into integral reduction, which we
used in our tests.  These showcase a significant increase in performance
compared to traditional methods.  Further improvements might be achieved via a
more refined strategy and additional optimizations, which we leave to future
work.

In general, given an integral family $F$, TI identities map some of its sectors
into sectors of new, simpler families $F_i$.  As already stated, we refer to
these as TI-mapped sectors and TI families respectively.  In this context, four
types of identities can be considered.
\begin{enumerate}
    \item Identities between integrals of the original family $F$.
    \item Transverse integration (TI) identities which map integrals of some sectors of $F$ into the TI families $F_i$.
    \item Identities between integrals belonging to the same TI family $F_i$.  These can be simply a new Laporta-like system of identities for $F_i$ or, by exploiting its simplicity with respect to the one for $F$, its full analytic solution might be reconstructed and used instead.
    \item Identities relating integrals of different TI families $F_i$ and $F_j$, with $i\neq j$.
\end{enumerate}
Finding the optimal combination of these identities for the reduction of a
given family is still an open problem, but in the following we will show one  which yields significant performance improvements in a number of cutting-edge applications.

\subsection{General strategy}
\label{sec:strategy}

Here we test a strategy that represents a good compromise between ease of
implementation and performance benefits.  Further improvements in performance
may be achieved by a more complex strategy and by further optimizing some of
its steps.

In the tests described below, we use TI identities for sectors with fewer legs
only, following section~\ref{sec:fewerlegs}.  TI identities for factorizable
sectors (described in section~\ref{sec:factorizable}) will not be considered
here, since they need a significantly more complex implementation in the
general case, being a sequence of transverse integrations.  Identities for
factorizable sectors have been tested with ad-hoc implementations for some
simple families, such as the one described in~\ref{sec:factorizableexample},
but a generic implementation and its impact on performance for more complex
applications is left to future work.

Consider an integral family $F$ with $e$ independent external legs.  We decide
a priori on a lower number $e'$ of external legs such that $e'<e$ and the reduction
for the TI families $F_i$ with $e'$ legs are simple enough to be reconstructed
\emph{analytically}\footnote{We make an exception to this rule for the \texttt{ttH} family, as described below.} in a relatively short time (in the order of minutes on a
modern machine).  We also restrict the list of TI-mapped sectors to be a subset of the \emph{unique sectors} of $F$, i.e.\ of those that cannot be mapped to
other sectors of $F$.

Among the unique sectors of $F$, we identify a minimal subset $\mathcal{G}_{TI}$ of
\emph{generating TI-mapped sectors}, namely a subset of sectors such that
\begin{itemize}
    \item sectors in $\mathcal{G}_{TI}$ have $e'$ or fewer external legs,
    \item every other unique sector of $F$ with $e'$ or fewer legs is a subsector of at least one sector of $\mathcal{G}_{TI}$.
\end{itemize}

For each sector in $S_{F,\vec{a}}\in\mathcal{G}_{TI}$, we define a new
TI integral family $F_i$.  The proper denominators of the top sector
of $F_i$ are identified with the proper denominators of the sector
$S_{F,\vec{a}}$, up to the shift in the loop momenta described in
eq.~\eqref{eq:kshift}.  We complete the list of proper denominators by adding a list of independent ISPs, which depends on the TI family $F_i$.  Following
section~\ref{sec:fewerlegs} we are thus able to build TI identities
mapping all integrals belonging to each generating TI-mapped sector $S_{F,\vec{a}}$ and its subsectors
to integrals in the respective TI family $F_i$.

We now turn to the choice of identities that we use for the reduction to master
integrals.  For definiteness, we consider the reduction to master integrals of
all integrals of family $F$ up to a maximum rank $s=s_{\textrm{max}}$ and up to
a maximum number of dots $u=u_{\textrm{max}}$.
In most cases, using seed integrals with $s\leq s_{\textrm{max}}$ and $u\leq
u_{\textrm{max}}$ is sufficient in a Laporta reduction, although sometimes
increasing $s_{\textrm{max}}$ or $u_{\textrm{max}}$ in the seed integrals is
required.  Using seeds with $u\leq u_{\textrm{max}}+1$ for sector
mapping is
also typically convenient.  Our goal is simplifying this system of identities
by using a lower number of equations, to be complemented by suitable TI
identities and integral identities satisfied by the TI families $F_i$.

Recalling the four types of identities defined above, in our tests we proceed
as follows.
\begin{enumerate}
    \item We generate a simpler Laporta system for $F$.  For sectors that are not TI-mapped, we choose the same seed integrals we would in a traditional Laporta reduction, i.e.\ with $s\leq s_{\textrm{max}}$ and $u\leq u_{\textrm{max}}$.  We also add to these some identities for TI-mapped sectors which, while not strictly necessary, reduce the number of integral reductions needed for the TI families $F_i$, without increasing too much the complexity of the system.  More precisely, consider a TI-mapped sector $S_{F,\vec{a}}$ with $t$ proper denominators.  If $s'$ is the maximum $s$ appearing  in the seed integrals of the parent sectors of $S_{F,\vec{a}}$ with $t+1$ proper denominators, then seed integrals for $S_{F,\vec{a}}$ will be chosen with $s\leq s'-1$ (as an exception, if $s'\leq 1$, we choose seeds with $s\leq 1$ for $S_{F,\vec{a}}$).  In other words, we choose fewer seed integrals for sectors with fewer and fewer denominators.  Note that similar choices of seed integrals have already been used in other contexts~\cite{Driesse:2024xad, Badger:2023mgf, Guan:2024byi}.\footnote{We thank Johann Usovitsch for interesting discussions and suggestions on this point.}
    \item Feynman integrals which are independent with respect to the linear system of the previous point are either MIs of $F$ or integrals belonging to TI-mapped sectors.  For the latter, we use TI identities which map them to the simpler integral families $F_i$, following section~\ref{sec:fewerlegs} (see also section~\ref{sec:implementation} for more details).
    \item For each integral appearing on the r.h.s.\ of the TI identities, we consider its analytic reduction to master integrals within the corresponding TI family $F_i$.
    \item As explained in section~\ref{sec:mappings}, there might be relations between the MIs of different TI families $F_i$.  However, these are easily found and included in a (relatively short) list of identitities between the master integrals of different families $F_i$.  In particular, given their relative simplicity, the full analytic form of these mappings is found and used.
\end{enumerate}
By combining these identities we obtain a complete a reduction to independent
master integrals for family $F$.

We stress that the identities in points 2 to 4 are merely a list of
linear substitutions.  In other words, they combine into a
triangular linear system or, equivalently, in a sequence of matrix
multiplications.  Hence, only the identities in point 1 generate a
non-trivial system of equations, albeit significantly simpler than the
one we would generate without considering TI identities.

For the~\texttt{ttH} family defined below, a slight variation of the setup described above was tested.  In that example, we integrated out subsectors with $e'+1=4$ external legs, but we did \emph{not} reconstruct analytically the solutions of the IBP identities for the TI families, in point 3.  Instead, we generated a Laporta system for them and solved it numerically over finite fields.

While solving systems of equations, one needs to sort the unknowns by defining
a \emph{weight}, such that unknowns with higher weight are always expressed in
terms of unknowns with lower weight.  This choice also affects the performance
of the linear solver.  For the identities in point 1, as a first criterion we
assign a lower weight to integrals belonging to sectors with $e'$ or fewer
external legs.  As a second criterion, when the first one yields a tie, we use
the weight proposed in Appedix~B of~\cite{Peraro:2019svx}.

\subsection{Implementation}
\label{sec:implementation}

We now give more details on our implementation of the general strategy outlined
in the previous subsection.  We implemented the numerical reduction to master
integrals over finite fields, using the \textsc{Mathematica} interface of the
\textsc{FiniteFlow} program~\cite{Peraro:2019svx}.  The definition and analysis
of integral families, as well as the generation of integral identities, have been
done via a custom \textsc{Mathematica} package built on top of
\textsc{FiniteFlow}.

We first identify the list of independent external legs for each sector of $F$
and the corresponding shift in the loop momenta.  We achieve this with a simple
algebraic algorithm, that we describe in appendix~\ref{sec:ffextlegs}.  From
this, we easily identify the list of generating TI-mapped sectors
$\mathcal{G}_{TI}$ via a top-down approach.

Identities among integrals of family $F$ are generated using traditional
methods, as in the Laporta algorithm, which we reviewed in
section~\ref{sec:background}, except that we generate fewer seed integrals,
hence fewer equations, as described in section~\ref{sec:strategy}.

We now turn to the implementation of TI identities.  In principle, we could
simply follow section~\ref{sec:fewerlegs} to find the identity we need for each
integral belonging to TI-mapped sectors that still needs to be reduced after
solving the first system of equations.  By combining those steps analytically,
however, one generally obtains huge expressions for these identities.  We work
around this by combining two simple tricks.
The first consists in splitting TI
identities into three much simpler steps, namely
\begin{enumerate}
    \item identities implementing the substitution in eq.~\eqref{eq:denFtodennewFort}, which rewrites the generalized denominators of $F$ in terms of the denominators of the TI family $F_i$ and orthogonal scalar products, yielding a combination of integrals in the form of eq.~\eqref{eq:Ikiortpj},
    \item identities implementing the tensor decomposition in the transverse space, rewriting any integral of the form in eq.~\eqref{eq:Ikiortpj} obtained in the previous step as a linear combination of integrals in the form of eq.~\eqref{eq:Ikiortkj},
    \item identities implementing the substitution in eq.~\eqref{eq:kiortkjorttopar}, with the r.h.s.\ replaced by a combination of generalized denominators of $F_i$.
\end{enumerate}
These three identities will thus be combined numerically over finite fields.
While this strongly reduces the size of the expressions, it doesn't completely
fix the issue for complex multi-scale integrals having high rank.  This happens
in particular for step 1 above, and sometimes also for step 3. We note, however,
that the identities in eq.~\eqref{eq:denFtodennewFort} and
eq.~\eqref{eq:Ikiortkj} for individual generalized denominators or scalar
products are relatively simple.  Moreover, they are simply substitutions of
loop variables which we can implement at the integrand level.  When these are
raised to high powers or multiplied to become high-rank integrals, on the other
hand, expressions can still become very large.
Finite-field technologies, however, allow us to fix this issue with our second trick. We first encode the
expressions for the required generalized denominators and scalar products as
(linear) polynomials in the loop variables appearing on the r.h.s.\ of the
equations we use in steps 1 and 3 above. The coefficients of such polynomials, which only depend on the
external variables, are instead treated numerically over finite fields.  This
drastically simplifies their expressions.  Hence, for each numerical
evaluation, we can afford to build higher powers of (products of) these
polynomials via a sequence of polynomial multiplications over finite fields.
At the time of writing, \textsc{FiniteFlow} does not expose polynomial
multiplication algorithms to its \textsc{Mathematica} interface, so these have
been implemented as a sequence of linear substitutions, for which
\textsc{FiniteFlow}'s sparse solver was used.
A more specialized and optimized algorithm might be used in the future, but
this was sufficient for our applications.  These two tricks thus allowed us to
implement complex TI identities without ever dealing with huge expressions.

Analytic IBP solutions for the TI families $F_i$ are simply implemented by
evaluating the coefficients of the reductions as rational functions of the
invariants.  The only optimization we used in this step is avoiding the
evaluation of duplicate coefficients appearing in multiple places.  More
aggressive optimizations strategies, such as identifying common subexpressions
or factors, may be considered in the future.

TI identities (split into three steps, as described before) and
solutions of IBP identities for the TI families need to be combined.  We accomplish this via a sequence of sparse matrix multiplications.\footnote{Each of these matrix multiplications is implemented using the \texttt{FFAlgTakeAndAddBL} algorithm of \textsc{FiniteFlow}, since it is more flexible than dedicated matrix multiplication algorithms and allows, in particular, for a sparse representation of the output.}  Finally, the latter is combined with the solution of the
simplified linear system for $F$ via a sparse matrix multiplication of their
respective coefficients, obtaining the final reduction.

These steps implement a numerical algorithm for the reduction of the considered
set of integrals in $F$.  We exploit the features of \textsc{FiniteFlow} which allow to define new algorithms by combining more basic ones as building blocks,  using dataflow graphs via a high-level interface (we use the \textsc{Mathematica} one).

\begin{figure}[h]
    \centering
    \includegraphics[width=.9\linewidth]{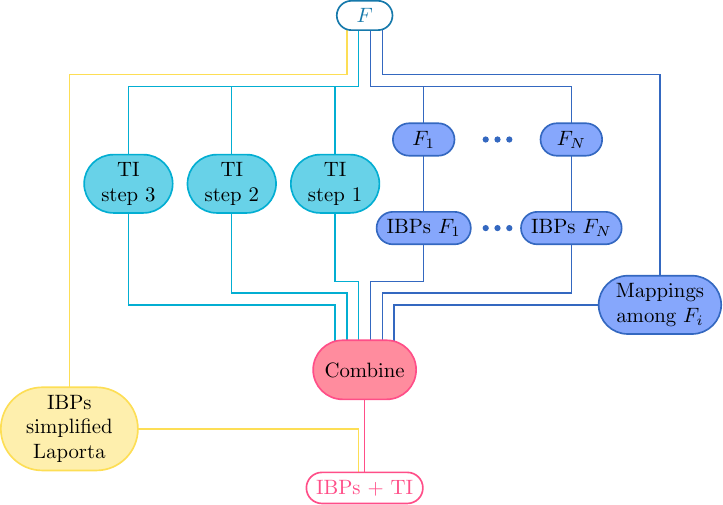}
    \caption{Schematic dataflow graph of our implementation of IBP reduction combined with TI identities.}
    \label{fig:flowchart}
\end{figure}

A schematic picture of a dataflow graph defining our algorithm is shown in figure~\ref{fig:flowchart}.  The input node $F$, which represents the input variables of our calculation, is a list which contains the space-time dimension $D$ and the invariants that describe the integral family $F$.  The nodes named $F_i$ compute instead the invariants for the simpler TI families $F_i$ in terms of the ones of $F$.  These are the inputs of the IBP reductions of each TI family $F_i$.  The latter are thus combined with the other steps of the calculation -- namely, the TI identities split into three steps, the simplified Laporta system for family $F$ and mappings between different TI families $F_i$ -- to yield a reduction for family $F$.

\pagebreak
\subsection{Examples}
\label{sec:examples}
We now show several examples of the transverse integration method
applied to reduction of two-loop Feynman integral families.

In particular, we report on the relative performance improvement of a
numerical evaluation over finite fields of the reduction, with respect
to an equivalent one that uses a more traditional Laporta algorithm.
The latter has also been implemented using the choice of seed
integrals proposed in~\cite{Peraro:2019svx} (also described in
section~\ref{sec:strategy}) and the weight choice proposed in
Appendix~B of the same reference.  We stress that the choice of seed
integrals for our new approach and the traditional approach we compare
against is thus identical, except for sectors where simplifications
are enabled by the use of TI identities. Both approaches have been
implemented using the \textsc{Mathematica} interface of the
\textsc{FiniteFlow} program.

For all integral families we consider, we reduce the full set of
integrals belonging to non-zero sectors with $u=0$ and
$s\leq s_{\textrm{max}}$ for some maximum rank $s_{\textrm{max}}$
specified below. In the following, we also break down the percentages of time spent in each main step of our method (note that they generally don't add up to $100\%$ since additional time is needed to combine the various types of identities).

The explicit definition of the TI families defined for the reductions, as well as diagrams representing them, are instead presented in appendix~\ref{app:TIfams}.

\subsubsection*{Double box with one external mass (\texttt{db})}
The double box family with one external mass, pictorially represented in
figure~\ref{fig:dbox1m}, was already presented in section~\ref{sec:pedagogical}.
We stress that this example is mainly presented because of its simplicity, but since its IBP reduction is already very easy using traditional methods, it should not be regarded as a significant target for performance improvements.

The generalized denominators of the family are summarized in eq.~\eqref{eq:db_props}.
We choose generating sectors with $e'+1 = 3$ or fewer
external legs and we find $4$ generating TI-mapped sectors $\mathcal{G}_{TI}$.
We map them into new TI families.
We reduce integrals
of this family up to rank $4$. With this setup we observe a performance
improvement of a factor $1.6$ for the evaluation of the reduction over
finite-fields, compared to a traditional Laporta system.
The evaluation timings are roughly split as follows:
\begin{itemize}
    \item solving the simplified Laporta IBP system:
        $83 \%$,
    \item evaluating the coefficients of the TI identities:
        $8.5 \%$,
    \item evaluating the solution of the IBP system for the TI families:
        $2 \%$.
\end{itemize}

\subsubsection*{Massless pentabox (\texttt{pb})}
\label{ex:PB}
\begin{figure}
    \centering
    \includegraphicsbox[]{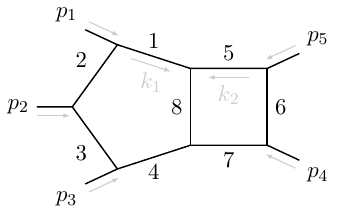}
    \caption{\texttt{pb} family.}
    \label{fig:pbox}
\end{figure}
The massless pentabox family, pictorially represented in figure~\ref{fig:pbox},
is described by the following kinematics
\begin{alignat}{3}
    &s_{12} = (p_1+p_2)^2,\quad
    &
    &s_{23} = (p_2+p_3)^2,\quad
    &
    &s_{34} = (p_3+p_4)^2,\nonumber\\
    &s_{45} = (p_4+p_5)^2,\quad
    &
    &s_{51} = (p_5+p_1)^2,\quad
    &
    &p_1^2 = p_2^2 = p_3^2 = p_4^2 = p_5^2 = 0,
\end{alignat}
and the external momenta satisfy $\sum_{i=1}^5 p_i =0$.

The defining list of generalized denominators for this family is given by
\begin{alignat}{3}
    &D_{\texttt{pb},1} ={} k_1^2,\quad
    &
    &D_{\texttt{pb},2} ={} \brk{k_1-p_1}^2,\quad
    &
    &D_{\texttt{pb},3} ={} \brk{k_1-p_1-p_2}^2,
    \nonumber \\
    &D_{\texttt{pb},4} ={} \brk{k_1-p_1-p_2-p_3}^2,\quad
    &
    &D_{\texttt{pb},5} ={} k_2^2,\quad
    &
    &D_{\texttt{pb},6} ={} \brk{k_2-p_5}^2,
    \nonumber \\
    &D_{\texttt{pb},7} ={} \brk{k_2+p_1+p_2+p_3}^2,\quad
    &
    &D_{\texttt{pb},8} ={} \brk{k_1+k_2}^2,\quad
    &
    &D_{\texttt{pb},9} ={} \brk{k_1+p_5}^2,
    \nonumber \\
    &D_{\texttt{pb},10} ={} \brk{k_2+p_1}^2,\quad
    &
    &D_{\texttt{pb},11} ={} \brk{k_2+p_1+p_2}^2.
\end{alignat}
We apply our algorithm for transverse integration choosing generating sectors
with $e'+1=4$ or fewer external legs. We find $5$ generating TI-mapped sectors $\mathcal{G}_{TI}$, which define the corresponding
 new TI families.
We reduce integrals of this family up to rank $5$ and observe a performance improvement of
a factor $4.0$ for the evaluation of the reduction over finite-fields, compared
to a traditional Laporta system.
The total runtime is roughly split as follows:
\begin{itemize}
    \item solving the simplified Laporta IBP system:
        $73 \%$,
    \item evaluating the coefficients of the TI identities:
        $7.5 \%$,
    \item evaluating the solution of the IBP system for the TI families:
        $12 \%$.
\end{itemize}
\subsubsection*{Massless non-planar double pentagon (\texttt{dp})}
\label{ex:DP}
\begin{figure}
    \centering \includegraphicsbox[]{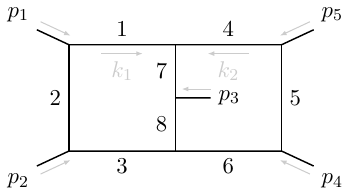}
    \caption{\texttt{dp} family.}
    \label{fig:dp}
\end{figure}
The non-planar two-loop five-point double pentagon family, pictorially represented in figure~\ref{fig:dp}, is described by the following kinematics,
\begin{alignat}{3}
    &s_{12} = (p_1+p_2)^2,\qquad
    &
    &s_{23} = (p_2+p_3)^2,\qquad
    &
    &s_{34} = (p_3+p_4)^2,\nonumber\\
    &s_{45} = (p_4+p_5)^2,\qquad
    &
    &s_{51} = (p_5+p_1)^2,\qquad
    &
    &p_1^2 = p_2^2 = p_3^2 = p_4^2 = p_5^2 = 0,
\end{alignat}
and the external legs satisfy $\sum_{i=1}^5 p_i = 0$.

Its set of generalized denominators reads
\begin{alignat}{3}
    &D_{\texttt{dp},1} ={} k_1^2, \quad
    &
    &D_{\texttt{dp},2} ={} \brk{k_1-p_1}^2, \quad
    &
    &D_{\texttt{dp},3} ={} \brk{k_1-p_1-p_2}^2,
    \nonumber \\
    &D_{\texttt{dp},4} ={} k_2^2, \quad
    &
    &D_{\texttt{dp},5} ={} \brk{k_2+p_1+p_2+p_3+p_4}^2, \quad
    &
    &D_{\texttt{dp},6} ={} \brk{k_2+p_1+p_2+p_3}^2,
    \nonumber \\
    &D_{\texttt{dp},7} ={} \brk{k_1+k_2}^2, \quad
    &
    &D_{\texttt{dp},8} ={} \brk{k_1+k_2+p_3}^2, \quad
    &
    &D_{\texttt{dp},9} ={} \brk{k_1+p_5}^2,
    \nonumber \\
    &D_{\texttt{dp},10} ={} \brk{k_2+p_1}^2, \quad
    &
    &D_{\texttt{dp},11} ={} \brk{k_2+p_1+p_2}^2.
\end{alignat}
We apply our algorithm for transverse integration choosing the generating
sectors with $e'+1 = 4$ external legs. We find $8$ generating TI-mapped sectors,
mapped into new TI families. We reduce integrals
of this family up to rank $5$. With this setup we observe a performance
improvement of a factor $3.3$ for the evaluation of the reduction over
finite-fields, compared to a traditional Laporta system.
The evaluation timings are roughly split as follows:
\begin{itemize}
    \item solving the simplified Laporta IBP system:
        $70 \%$,
    \item evaluating the coefficients of the TI identities:
        $21 \%$,
    \item evaluating the solution of the IBP system for the TI families:
        $6 \%$.
\end{itemize}

\subsubsection*{Pentabox of top-pair plus Higgs production (\texttt{ttH})}
\label{ex:ttH}
\begin{figure}
    \centering
    \includegraphics[]{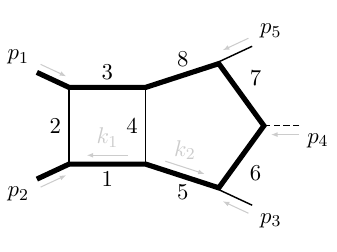}
    \caption{\texttt{ttH} family.  Thick external and internal lines are massive, with mass $m_{\texttt{t}}$, while the dashed external line has mass $m_{\texttt{H}}$.}
    \label{fig:ttH}
\end{figure}
In this example we consider a two-loop pentabox family contributing to Higgs production in association with a top pair in hadron collisions.  The family, pictorially represented in figure~\ref{fig:ttH}, is described by the following kinematics,
\begin{alignat}{3}
    &s_{12} = (p_1+p_2)^2,\qquad
    &
    &s_{23} = (p_2+p_3)^2,\qquad
    &
    &s_{34} = (p_3+p_4)^2,\nonumber\\
    &s_{45} = (p_4+p_5)^2,\qquad
    &
    &s_{51} = (p_5+p_1)^2,\qquad
    &
    &p_1^2 = p_2^2 = m_{\texttt{t}}^2 \nonumber\\
    &p_3^2 = p_5^2 = 0,\qquad
    &
    &p_4^2 = m_{\texttt{H}}^2.
\end{alignat}
Its external legs satisfy $\sum_{i=1}^5 p_i = 0$.

The list of generalized denominators is chosen as follows
\begin{alignat}{2}
    &D_{\texttt{ttH},1} ={} k_1^2-m_{\texttt{t}}^2, \quad
    &
    &D_{\texttt{ttH},2} ={} \brk{k_1+p_2}^2, \nonumber\\
    &D_{\texttt{ttH},3} ={} \brk{k_1+p_1+p_2}^2-m_{\texttt{t}}^2, \quad
    &
    &D_{\texttt{ttH},4} ={} \brk{k_1+k_2}^2, \nonumber\\
    &D_{\texttt{ttH},5} ={} k_2^2 - m_{\texttt{t}}^2, \quad
    &
    &D_{\texttt{ttH},6} ={} \brk{k_2+p_3}^2-m_{\texttt{t}}^2,\nonumber\\
    &D_{\texttt{ttH},7} ={} \brk{k_2+p_3+p_4}^2-m_{\texttt{t}}^2, \quad
    &
    &D_{\texttt{ttH},8} ={} \brk{k_2+p_3+p_4+p_5}^2-m_{\texttt{t}}^2, \nonumber\\
    &D_{\texttt{ttH},9} ={} \brk{k_1+p_5}^2, \quad
    &
    &D_{\texttt{ttH},10} ={} \brk{k_2+p_2}^2, \nonumber\\
    &D_{\texttt{ttH},11} ={} \brk{k_1+p_4+p_5}^2.
\end{alignat}
We apply our algorithm for transverse integration choosing the generating sectors with $e'+1 = 4$ or fewer external legs. We find $8$ generating TI-mapped sectors, mapped into new TI families. We reduce integrals of this family up to rank $5$.

As briefly mentioned before, for this family we tested a variation of our the setup, where the reductions for the TI families were not reconstructed analytically but only evaluated numerically from a Laporta system.

With this setup, we observe a performance improvement of a factor $2.3$ for the evaluation of the reduction over finite-fields, compared to a traditional Laporta system. The timings for evaluation is roughly split as follows:
\begin{itemize}
\item evaluating the IBP system for the the simplified Laporta system: $48 \%$,
\item evaluating the coefficients of the TI identities: $22 \%$,
\item solving of the IBP systems for the TI families: $23 \%$.\end{itemize}

\subsubsection*{Massless two-loop six-point ladybug (\texttt{lb})}
\label{ex:LB}
\begin{figure}
    \centering
    \includegraphicsbox{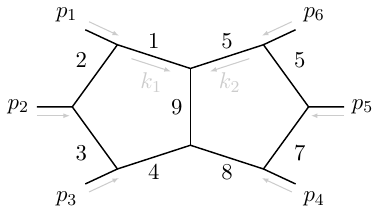}
    \caption{\texttt{lb} family.}
    \label{fig:lb}
\end{figure}
The two-loop six-point ladybug family, pictorially represented in figure~\ref{fig:lb}, is described by the following kinematics,
\begin{alignat}{3}
    &s_{12} = (p_1+p_2)^2,\qquad
    &
    &s_{23} = (p_2+p_3)^2,\qquad
    &
    &s_{34} = (p_3+p_4)^2,\nonumber\\
    &s_{45} = (p_4+p_5)^2,\qquad
    &
    &s_{56} = (p_5+p_1)^2,\qquad
    &
    &s_{61} = (p_6+p_1)^2,\nonumber\\
    &s_{123} = (p_1+p_2+p_3)^2,\qquad
    &
    &s_{234} = (p_2+p_3+p_4)^2,\qquad
    &
    &s_{345} = (p_3+p_4+p_5)^2,\nonumber\\
    &p_1^2 = p_2^2 = p_3^2 = p_4^2 = p_5^2 =p_6^2 =0.
\end{alignat}
Its external legs satisfy $\sum_{i=1}^6 p_i = 0$.  Note that we take the external momenta to be $D$-dimensional, hence there are 9 independent kinematic invariants and five of the six external momenta are linearly independent.

The list of its generalized denominators is
\begin{alignat}{3}
    &D_{\texttt{lb},1} ={} k_1^2, \quad
    &
    &D_{\texttt{lb},2} ={} \brk{k_1-p_1}^2, \quad
    &
    &D_{\texttt{lb},3} ={} \brk{k_1-p_1-p_2}^2,
    \nonumber \\
    &D_{\texttt{lb},4} ={} \brk{k_1-p_1-p_2-p_3}^2, \quad
    &
    &D_{\texttt{lb},5} ={} k_2^2, \quad
    &
    &D_{\texttt{lb},6} ={} \brk{k_2-p_6}^2,
    \nonumber \\
    &D_{\texttt{lb},7} ={} \brk{k_2-p_5-p_6}^2, \quad
    &
    &D_{\texttt{lb},8} ={} \brk{k_2-p_4-p_5-p_6}^2, \quad
    &
    &D_{\texttt{lb},9} ={} \brk{k_1+k_2}^2,
    \nonumber \\
    &D_{\texttt{lb},10} ={} \brk{k_1+p_6}^2, \quad
    &
    &D_{\texttt{lb},11} ={} \brk{k_1+p_5+p_6}^2, \quad
    &
    &D_{\texttt{lb},12} ={} \brk{k_2+p_1}^2, \quad
    \nonumber \\
    &D_{\texttt{lb},13} ={} \brk{k_2+p_1+p_2}^2. \quad
\end{alignat}
We apply our algorithm for transverse integration, choosing generating
sectors with $e'+1 = 4$ or fewer external legs. We find $19$ generating TI-mapped sectors,
mapped into simpler TI families. We reduce integrals
of this family up to rank $5$. With this setup, we observe a performance
improvement of a factor $2.7$ for the evaluation of the reduction over
finite-fields, compared to a traditional Laporta system.
The evaluation timings are roughly split as follows:
\begin{itemize}
    \item solving the simplified Laporta IBP system:
        $58 \%$,
    \item evaluating the coefficients of the TI identities:
        $29 \%$,
    \item evaluating the solution of the IBP system for the TI families:
        $9 \%$.
\end{itemize}

\section{Conclusions and outlook}
\label{sec:conclusions}
In this paper, we provide a formulation of transverse integration (TI) identities via tensor decomposition and we apply them to the reduction of Feynman integrals to master integrals. Despite being already used in other contexts, the application of TI to traditional IBP reduction was still unexplored. We present a general algorithm to obtain the TI identities for sectors of a given family that can be mapped to new families -- which we call TI families -- having either fewer independent external legs or that can be factored into products of lower-loop integrals. We describe a proof-of-concept implementation of these identities in the context of the Laporta algorithm. More in detail, we implement a numerical algorithm for IBP reduction over finite fields using the~\textsc{FiniteFlow} program, which includes a combination of TI and IBP identities. We prove that the use of TI identities combined with the Laporta algorithm provides a significant computational advantage on several cutting-edge examples.

In future works, we plan to add further optimizations, such as combining TI identities with syzygy techniques, automating them also for the so-called factorizable sectors (which we did not include in our benchmarks, since their implementation is more involved, but could provide sizable performance benefits) and using specialized algorithms for some steps of the implementation, such as polynomial multiplication. Moreover, our algorithm may also be also used recursively to find (analytic or numeric) solutions for the TI families themselves. Given the improvements that this technique can bring to state-of-the-art IBP reduction, the release of a public implementation of it is also within our plans.

\acknowledgments
We thank Pierpaolo Mastrolia, Kay Sch\"{o}nwald and Simone Zoia for comments and feedback on the draft and Johann Usovitsch for discussions and suggestions on the seeding for IBP identities.
VC and TP received funding from the European Research Council (ERC) under the European Union’s Horizon Europe research and innovation programme grant agreement 101040760 (ERC Starting Grant FFHiggsTop), GF received funding from the European Research Council (ERC) under the European Union’s Horizon 2020 research and innovation programme grant agreement 101019620 (ERC Advanced Grant TOPUP) and from the UZH Candoc scheme (Candoc grant Nifty LooPS).
%This work received funding from the European Research Council (ERC) under the European Union’s Horizon 2020 research and innovation programme grant agreement 101019620 (ERC Advanced Grant TOPUP) and under the European Union’s Horizon Europe research and innovation programme grant agreement 101040760 (ERC Starting Grant FFHiggsTop).

\appendix

\section{Identification of TI-mapped sectors and loop shifts}
\label{app:algsTI}
Here we describe two simple algorithms useful for detecting and processing Feynman
integrals amenable to transverse integration.  In particular, we identify the external legs of a sector and determine whether a sector is factorizable, with its list of factors.

Both algorithms make use of simple linear algebra, starting from the expression of the momenta $l_j$ which appear in the definition of the \emph{proper denominators}, which have the form in eq.~\eqref{eq:gen_den_1}.  As already stated, these are linear combinations of external and loop momenta ($p_j$ and $k_j$ respectively), as follows
\begin{equation}\label{eq:lj}
  l_i = \sum_{j=1}^\ell \alpha_{ij}\, k_j + \sum_{j=1}^e\beta_{ij}\, p_j,
\end{equation}
with $\alpha_{ij},\beta_{ij}\in\{0,\pm 1\}$.
\subsection{Identifying external legs of a sector}
\label{sec:ffextlegs}
To identify the independent external momenta of a given sector, we start from the list of the momenta $l_j$ running in its proper denominators, expressed in terms of loop momenta and independent external momenta as in eq.~\eqref{eq:lj}, and we build the linear system
\begin{equation}\label{eq:ljffextlegs}
  \sum_{j=1}^\ell \alpha_{ij}\, k_j + \sum_{j=1}^e\beta_{ij}\, p_j = 0.
\end{equation}
In other words we set $l_j=0$ and solve for the variables $k_j$ and $p_j$, assigning higher weight to (i.e.\ eliminating with higher priority) the loop momenta $k_j$.  The solution of this system writes all loop momenta and a subset of the external momenta as linear combinations of the other external momenta, namely
\begin{equation}\label{eq:ffextlegssol}
  k_i = \sum_{j} a_{ij}\, p_j, \qquad p_i = \sum_{j} b_{ij}\, p_j,
\end{equation}
where the sums run over the external momenta which have not been fixed by the system of equations and  $a_{ij},b_{ij}\in\{0,\pm 1\}$.  From this solution we read the $D_\parallel$ independent external momenta $q_i$ of the sector as the non-trivial linear combination of momenta of the form
\begin{equation}\label{eq:qisol}
  q_i^\mu = p_i^\mu - \sum_{j} b_{ij}\, p_j^\mu.
\end{equation}
The shift of the loop momenta which makes the dependence on $q_i$ manifest in the proper denominator expressions (cfr.\ eq.~\eqref{eq:kshift}) is
\begin{equation}
  k_i^\mu \to k_{s,i}^\mu = k_i^\mu + \sum_{j} a_{ij}\, p_j^\mu.
\end{equation}
The rationale is the following.  The linear combinations in eq.~\eqref{eq:qisol} vanish on the solutions of the system, which implies that -- as one would see if we hadn't set $l_j=0$ -- they are linear combinations of the momenta $l_j^\mu$ running in the loop denominators.  Moreover, since we eliminated the loop momenta $k_j$ first in the solution of the system, these linear combinations are also independent of the loop momenta.  Hence, the non-trivial combinations in~\eqref{eq:qisol} are independent linear combinations of the momenta of the loop denominators which do not depend on the loop momenta, implying they are the independent linear combinations of the external momenta of the diagram that identifies the sector.

\subsection{Identifying factorizable sectors}
\label{sec:ffloopfactors}

To deduce if a given sector can be decomposed into a product of independent
lower-loop families, we again analyze the system~\eqref{eq:lj},
which in this case we may strip from the external momenta $p_i$, since the latter are irrelevant for this purpose.  We thus build from eq.~\eqref{eq:lj} the following linear system
\begin{align}
    l_i = \sum_{j=1}^\ell \alpha_{ij}\, k_j.
\end{align}
We solve this system with respect to the unkowns $k_j$ and $l_j$, assigning higher weight to the loop momenta $k_j$,\ i.e.\ eliminating the loop momenta with higher priority in the solution.  The solution expresses the loop momenta $k_j$ and some of the denominator momenta $l_j$ as linear combinations of the remaining $l_j$, namely
\begin{equation}\label{eq:ffloopfactorssol}
  k_i = \sum_{j} c_{ij}\, l_j, \qquad l_i = \sum_{j} d_{ij}\, l_j,
\end{equation}
where the sums run over the momenta $l_j$ that have not been fixed by the solution of the system and $c_{ij},d_{ij}\in \{0,\pm 1\}$.
We now promote the independent $l_j$, appearing in the r.h.s.\ of~\eqref{eq:ffloopfactorssol}, to be the new loop momenta and derive the corresponding
shift (cfr.\ with eq.~\eqref{eq:kshift})
\begin{equation}
  k_i^\mu \to k_{s,i}^\mu = \sum_{j} c_{ij}\, k_j^\mu.
\end{equation}
After this change of integration variables, any factorization of the loop becomes manifest.  Hence, at this point we simply identify subsets of loop momenta that partition the loop denominators, after the shift, into a subset which exclusively depends on them and a complementary subset that is independent of them.

\section{Definition of TI families in the examples}
\label{app:TIfams}
{\allowdisplaybreaks
  We give a detailed account of the TI families for each integral family presented in the examples of section~\ref{sec:examples}.  For each TI family, we report the list of $D_\parallel$ independent external momenta $q_i$, the shift of the loop momenta which makes the dependence of the proper denominators on $q_i$ manifest (see eq.~\eqref{eq:kshift}) and the list of generalized denominators of the TI family.  We use the convention that the proper denominators of the top sector of each TI family have the quadratic form in eq.~\eqref{eq:gen_den_2}, while its ISPs have the bi-linear form in eq.~\eqref{eq:gen_den_1}.  We also provide pictures of the Feynman diagrams which correspond to (the top sector of) each TI family.

\subsection*{Double box with one external mass}
The TI families are pictorially represented in figure~\ref{fig:dbox1m_new_families}.
\begin{figure}
    \centering
    \begin{subfigure}{.4\textwidth}
        \centering
        \includegraphicsbox{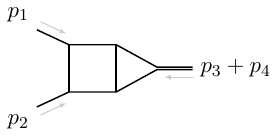}
        \caption{TI family \texttt{db1}. Same as \texttt{bt} in figure~\ref{fig:bt}.}
    \end{subfigure}
    \begin{subfigure}{.4\textwidth}
        \centering
        \includegraphicsbox{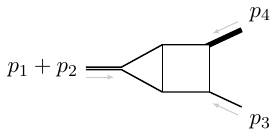}
        \caption{TI family \texttt{db2}.}
    \end{subfigure}
    \\
    \begin{subfigure}{.4\textwidth}
        \centering
        \includegraphicsbox{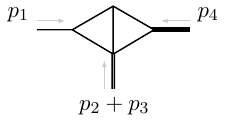}
        \caption{TI family \texttt{db3}.}
    \end{subfigure}
    \begin{subfigure}{.4\textwidth}
        \centering
        \includegraphicsbox{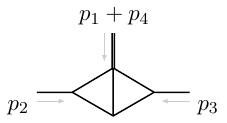}
        \caption{TI family \texttt{db4}.}
    \end{subfigure}
    \caption{
        \texttt{db} TI families.
    }
    \label{fig:dbox1m_new_families}
\end{figure}
The corresponding shifts of loop momenta read
\begin{align}
    \texttt{db1},\> \texttt{db2},\> \texttt{db3}&:
    \quad
    \brk{k_1,\> k_2} \mapsto \brk{k_1,\> k_2},
    \\
    \texttt{db4}&:
    \quad
    \brk{k_1,\> k_2} \mapsto \brk{k_1 - p_1,\> k_2 + p_1},
\end{align}
and the independent external momenta for each TI family are
\begin{align}
    \texttt{db1}&:
    \quad
    \brk{q_1,\> q_2} \defas \brk{p_1,\> p_2},
    \\
    \texttt{db2}&:
    \quad
    \brk{q_1,\> q_2} \defas \brk{p_1 + p_2,\> p_3},
    \\
    \texttt{db3}&:
    \quad
    \brk{q_1,\> q_2} \defas \brk{p_1,\> p_2 + p_3},
    \\
    \texttt{db4}&:
    \quad
    \brk{q_1,\> q_2} \defas \brk{p_2,\> p_3}.
\end{align}
The generalized denominators of the TI families are:
{\allowdisplaybreaks
\begin{alignat}{3}
    &D_{\texttt{db1}, 1} ={} k_1^2\quad,
    &
    &D_{\texttt{db1}, 2} ={} \brk{k_1+q_1}^2,
    \quad
    &
    &D_{\texttt{db1}, 3} ={} \brk{k_1+q_1+q_2}^2,
    \nonumber\\*
    &D_{\texttt{db1}, 4} ={} \brk{k_1+k_2}^2,
    \quad
    &
    &D_{\texttt{db1}, 5} ={} k_2^2,\quad
    &
    &D_{\texttt{db1}, 6} ={} \brk{k_2-q_1-q_2}^2,
    \nonumber\\*
    &D_{\texttt{db1}, 7} ={} k_2 \cdot q_2,
    \\ \nonumber\\
    &D_{\texttt{db2}, 1} ={} k_1^2,\quad
    &
    &D_{\texttt{db2}, 2} ={} \brk{k_1+q_1}^2,\quad
    &
    &D_{\texttt{db2}, 3} ={} \brk{k_1+k_2}^2,
    \nonumber\\*
    &D_{\texttt{db2}, 4} ={} k_2^2,
    \quad
    &
    &D_{\texttt{db2}, 5} ={} \brk{k_2-q_1-q_2}^2,
    \quad
    &
    &D_{\texttt{db2}, 6} ={} \brk{k_2-q_1}^2,\nonumber\\*
    &D_{\texttt{db2}, 7} ={} k_1 \cdot q_2,
\\ \nonumber\\
    &D_{\texttt{db3}, 1} ={} k_1^2,\quad
    &
    &D_{\texttt{db3}, 2} ={} \brk{k_1+q_1}^2,\quad
    &
    &D_{\texttt{db3}, 3} ={} \brk{k_1+k_2}^2,
    \nonumber\\*
    &D_{\texttt{db3}, 4} ={} k_2^2,
    \quad
    &
    &D_{\texttt{db3}, 5} ={} \brk{k_2-q_1-q_2}^2,
    \quad
    &
    &D_{\texttt{db3}, 6} ={} k_1 \cdot q_2,
    \nonumber\\*
    &D_{\texttt{db3}, 7} ={} k_2 \cdot q_2,
\\ \nonumber\\
    &D_{\texttt{db4}, 1} ={} k_1^2,\quad
    &
    &D_{\texttt{db4}, 2} ={} \brk{k_1+q_1}^2,\quad
    &
    &D_{\texttt{db4}, 3} ={} \brk{k_1+k_2}^2,
    \nonumber\\*
    &D_{\texttt{db4}, 4} ={} \brk{k_2-q_1-q_2}^2,
    \quad
    &
    &D_{\texttt{db4}, 5} ={} \brk{k_2-q_1}^2,
    \quad
    &
    &D_{\texttt{db4}, 6} ={} k_2 \cdot q_1,
    \nonumber \\*
    &D_{\texttt{db4}, 7} ={} k_1 \cdot q_2.
\end{alignat}
}

\subsection*{Massless pentabox}
\label{app:TIFamPB}
The TI families are pictorially represented in figure~\ref{fig:TIfam-pentabox}.
\begin{figure}
\centering
    \begin{subfigure}{.3\textwidth}
        \centering
        \includegraphicsbox{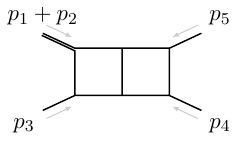}
        \caption{
            TI family \texttt{pb1}.
        }
    \end{subfigure}
    \begin{subfigure}{.3\textwidth}
        \centering
        \includegraphicsbox{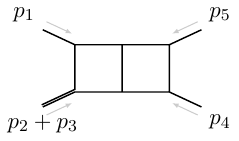}
        \caption{
            TI family \texttt{pb2}.
        }
    \end{subfigure}
    \\[.4cm]
    \begin{subfigure}{.3\textwidth}
        \centering
        \includegraphicsbox{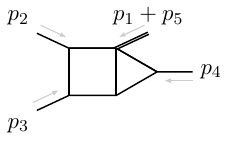}
        \caption{
            TI family \texttt{pb3}.
        }
    \end{subfigure}
    \begin{subfigure}{.3\textwidth}
        \centering
        \includegraphicsbox{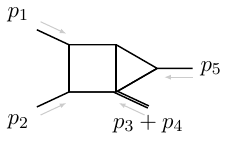}
        \caption{
           TI family \texttt{pb4}.
        }
    \end{subfigure}
    \begin{subfigure}{.3\textwidth}
        \centering
        \includegraphicsbox{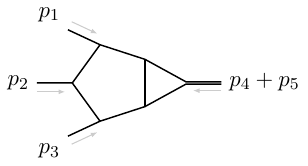}
        \caption{
           TI family \texttt{pb5}.
        }
    \end{subfigure}
    %\begin{subfigure}{.45\textwidth}
     %   \includegraphicsbox{figures/pb6}
     %   \caption{
     %       TI family \texttt{pb6}
     %   }
    %\end{subfigure}
    \caption{
        \texttt{pb} TI families.
    }
    \label{fig:TIfam-pentabox}
\end{figure}
The corresponding shifts of loop momenta read
\begin{align}
    \texttt{pb1},\> \texttt{pb2},\> \texttt{pb3},\> \texttt{pb4}&:
    \quad
    \brk{k_1,\> k_2} \mapsto \brk{k_1,\> k_2},
     \nonumber\\
    \texttt{pb5}&:
    \quad
    \brk{k_1,\> k_2} \mapsto \brk{k_1 - p_5,\> k_2 + p_5},
\end{align}
and the independent external momenta for each TI family are:
\begin{align}
    \texttt{pb1}&:
    \quad
    \brk{q_1,\> q_2,\> q_3} \defas \brk{p_1,\> p_2,\> p_3},
     \nonumber\\
    \texttt{pb2}&:
    \quad
    \brk{q_1,\> q_2,\> q_3} \defas \brk{p_1,\> p_2+p_3,\> p_5},
     \nonumber\\
    \texttt{pb3}&:
    \quad
    \brk{q_1,\> q_2,\> q_3} \defas \brk{p_1+p_2,\> p_3,\> p_5},
     \nonumber\\
    \texttt{pb4}&:
    \quad
    \brk{q_1,\> q_2,\> q_3} \defas \brk{p_1,\> p_2,\> p_5},
     \nonumber\\
    \texttt{pb5}&:
    \quad
    \brk{q_1,\> q_2,\> q_3} \defas \brk{p_2,\> p_3,\> p_1+p_5}.
\end{align}
The generalized denominators of the TI families are:
\begin{alignat}{3}
    &D_{\texttt{pb1}, 1} ={} k_1^2,\quad
    &
    &D_{\texttt{pb1}, 2} ={} \brk{k_1-q_1}^2,\quad
    &
    &D_{\texttt{pb1}, 3} ={} \brk{k_1-q_1-q_2}^2,
    \nonumber\\*
    &D_{\texttt{pb1}, 4} ={} \brk{k_1-q_1-q_2-q_3}^2,\quad
    &
    &D_{\texttt{pb1}, 5} ={} k_2^2,\quad
    &
    &D_{\texttt{pb1}, 6} ={} \brk{k_2+q_1+q_2+q_3}^2,
    \nonumber \\*
    &D_{\texttt{pb1}, 7} ={} \brk{k_1+k_2}^2,\quad
    &
    &D_{\texttt{pb1}, 8} ={} k_2 \cdot q_2,\quad
    &
    &D_{\texttt{pb1}, 9} ={} k_2 \cdot q_3,
\\ \nonumber\\
    &D_{\texttt{pb2}, 1} ={} k_1^2, \quad
    &
    &D_{\texttt{pb2}, 2} ={} \brk{k_1-q_1}^2, \quad
    &
    &D_{\texttt{pb2}, 3} ={} \brk{k_1-q_1-q_2}^2,
    \nonumber\\*
    &D_{\texttt{pb2}, 4} ={} k_2^2, \quad
    &
    &D_{\texttt{pb2}, 5} ={} \brk{k_2-q_3}^2, \quad
    &
    &D_{\texttt{pb2}, 6} ={} \brk{k_2+q_1+q_2}^2,
    \nonumber \\*
    &D_{\texttt{pb2}, 7} ={} \brk{k_1+k_2}^2,\quad
    &
    &D_{\texttt{pb2}, 8} ={} k_2 \cdot q_2,\quad
    &
    &D_{\texttt{pb2}, 9} ={} k_2 \cdot q_3,
\\ \nonumber\\
    &D_{\texttt{pb3}, 1} ={} k_1^2, \quad
    &
    &D_{\texttt{pb3}, 2} ={} \brk{k_1-q_1}^2, \quad
    &
    &D_{\texttt{pb3}, 3} ={} \brk{k_1-q_1-q_2}^2,
    \nonumber\\*
    &D_{\texttt{pb3}, 4} ={} k_2^2, \quad
    &
    &D_{\texttt{pb3}, 5} ={} \brk{k_2-q_3}^2,\quad
    &
    &D_{\texttt{pb3}, 6} ={} \brk{k_2+q_1+q_2}^2,
    \nonumber\\*
    &D_{\texttt{pb3}, 7} ={} \brk{k_1+k_2}^2,\quad
    &
    &D_{\texttt{pb3}, 8} ={} k_2 \cdot q_2,\quad
    &
    &D_{\texttt{pb3}, 9} ={} k_2 \cdot q_3,
\\ \nonumber\\
    &D_{\texttt{pb4}, 1} ={} k_1^2, \quad
    &
    &D_{\texttt{pb4}, 2} ={} \brk{k_1-q_1}^2, \quad
    &
    &D_{\texttt{pb4}, 3} ={} \brk{k_1-q_1-q_2}^2,
    \nonumber\\*
    &D_{\texttt{pb4}, 4} ={} k_2^2,\quad
    &
    &D_{\texttt{pb4}, 5} ={} \brk{k_2-q_3}^2,\quad
    &
    &D_{\texttt{pb4}, 6} ={} \brk{k_1+k_2}^2,
    \nonumber\\*
    &D_{\texttt{pb4}, 7} ={} k_2 \cdot q_1, \quad
    &
    &D_{\texttt{pb4}, 8} ={} k_2 \cdot q_2,\quad
    &
    &D_{\texttt{pb4}, 9} ={} k_1 \cdot q_3,
\\ \nonumber\\
    &D_{\texttt{pb5}, 1} ={} \brk{k_1-q_3}^2,\quad
    &
    &D_{\texttt{pb5}, 2} ={} \brk{k_1-q_1-q_3}^2, \quad
    &
    &D_{\texttt{pb5}, 3} ={} \brk{k_1-q_1-q_2-q_3}^2,
    \nonumber\\*
    &D_{\texttt{pb5}, 4} ={} k_2^2, \quad
    &
    &D_{\texttt{pb5}, 5} ={} \brk{k_2+q_1+q_2+q_3}^2, \quad
    &
    &D_{\texttt{pb5}, 6} ={} \brk{k_1+k_2}^2,
    \nonumber\\*
    &D_{\texttt{pb5}, 7} ={} k_2 \cdot q_2,\quad
    &
    &D_{\texttt{pb5}, 8} ={} k_1 \cdot q_3,\quad
    &
    &D_{\texttt{pb5}, 9} ={} k_2 \cdot q_3.
\end{alignat}
\subsection*{Massless non-planar double pentagon}
\label{app:TIFamDP}
The TI families are pictorially represented in figure~\ref{fig:tifam-dp}.
\begin{figure}
    \centering
    \begin{subfigure}{.4\textwidth}
        \centering
        \includegraphicsbox{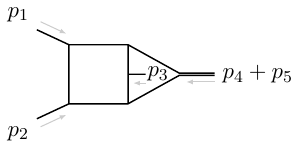}
        \caption{TI family \texttt{dp1}}
    \end{subfigure}
    \begin{subfigure}{.4\textwidth}
        \centering
        \includegraphicsbox{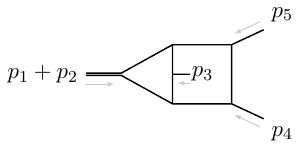}
        \caption{TI family \texttt{dp2}}
    \end{subfigure}
    \\[.4cm]
    \begin{subfigure}{.3\textwidth}
        \centering
        \includegraphicsbox{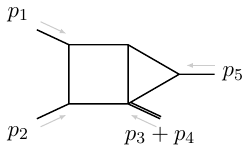}
        \caption{TI family \texttt{dp3}}
    \end{subfigure}
    \begin{subfigure}{.3\textwidth}
        \centering
        \includegraphicsbox{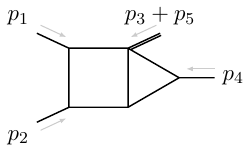}
        \caption{TI family \texttt{dp4}}
    \end{subfigure}
    \begin{subfigure}{.3\textwidth}
        \centering
        \includegraphicsbox{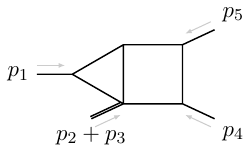}
        \caption{TI family \texttt{dp5}}
    \end{subfigure}
    \\[.4cm]
    \begin{subfigure}{.3\textwidth}
        \centering
        \includegraphicsbox{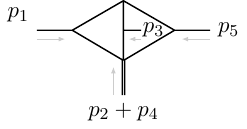}
        \caption{TI family \texttt{dp6}}
    \end{subfigure}
    \begin{subfigure}{.3\textwidth}
        \centering
        \includegraphicsbox{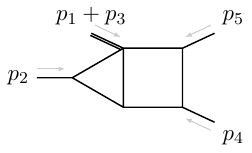}
        \caption{TI family \texttt{dp7}}
    \end{subfigure}
    \begin{subfigure}{.3\textwidth}
        \centering
        \includegraphicsbox{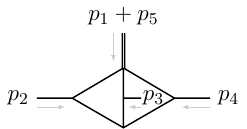}
        \caption{TI family \texttt{dp8}}
    \end{subfigure}

    \caption{\texttt{dp} TI families.}
    \label{fig:tifam-dp}
\end{figure}
The shift of loop momenta are as follows
\begin{align}
    \texttt{dp1},\> \texttt{dp2},\> \texttt{dp3},\> \texttt{dp5},\> \texttt{dp6}&:
    \quad
    \brk{k_1,\> k_2} \mapsto \brk{k_1,\> k_2},
     \nonumber\\
    \texttt{dp4}&:
    \quad
    \brk{k_1,\> k_2} \mapsto \brk{k_1,\> k_2 + p_5},
     \nonumber\\
    \texttt{dp7}&:
    \quad
    \brk{k_1,\> k_2} \mapsto \brk{k_1 - p_3,\> k_2 },
     \nonumber\\
    \texttt{dp8}&:
    \quad
    \brk{k_1,\> k_2} \mapsto \brk{k_1 - p_5,\> k_2 + p_5},
\end{align}
and the independent external momenta are:
\begin{align}
    \texttt{dp1}&:
    \quad
    \brk{q_1,\> q_2,\> q_3} \defas \brk{p_1,\> p_2,\> p_3},
     \nonumber\\
    \texttt{dp2}&:
    \quad
    \brk{q_1,\> q_2,\> q_3} \defas \brk{p_1+p_2,\> p_3,\> p_5},
     \nonumber\\
    \texttt{dp3}&:
    \quad
    \brk{q_1,\> q_2,\> q_3} \defas \brk{p_1,\> p_2,\> p_5},
     \nonumber\\
    \texttt{dp4}&:
    \quad
    \brk{q_1,\> q_2,\> q_3} \defas \brk{p_1,\> p_2,\> p_3+p_5},
     \nonumber\\
    \texttt{dp5}&:
    \quad
    \brk{q_1,\> q_2,\> q_3} \defas \brk{p_1,\> p_2+p_3,\> p_5},
     \nonumber\\
    \texttt{dp6}&:
    \quad
    \brk{q_1,\> q_2,\> q_3} \defas \brk{p_1,\> p_3,\> p_5},
     \nonumber\\
    \texttt{dp7}&:
    \quad
    \brk{q_1,\> q_2,\> q_3} \defas \brk{p_2,\> p_1+p_3,\> p_5},
     \nonumber\\
    \texttt{dp8}&:
    \quad
    \brk{q_1,\> q_2,\> q_3} \defas \brk{p_2,\> p_3,\> p_1+p_5}.
\end{align}
The generalized denominators of the TI families are:
\begin{alignat}{3}
    &D_{\texttt{dp1}, 1} ={} k_1^2,\quad
    &
    &D_{\texttt{dp1}, 2} ={} \brk{k_1-q_1}^2, \quad
    &
    &D_{\texttt{dp1}, 3} ={} \brk{k_1-q_1-q_2}^2,
    \nonumber\\*
    &D_{\texttt{dp1}, 4} ={} k_2^2, \quad
    &
    &D_{\texttt{dp1}, 5} ={} \brk{k_2+q_1+q_2+q_3}^2,\quad
    &
    &D_{\texttt{dp1}, 6} ={} \brk{k_1+k_2}^2,
    \nonumber\\*
    &D_{\texttt{dp1}, 7} ={} \brk{k_1+k_2+q_3}^2,\quad
    &
    &D_{\texttt{dp1}, 8} ={} k_2 \cdot q_2,\quad
    &
    &D_{\texttt{dp1}, 9} ={} k_2 \cdot q_3,
\\ \nonumber\\
    &D_{\texttt{dp2}, 1} ={} k_1^2,\quad
    &
    &D_{\texttt{dp2}, 2} ={} \brk{k_1-q_1}^2, \quad
    &
    &D_{\texttt{dp2}, 3} ={} k_2^2,
    \nonumber\\*
    &D_{\texttt{dp2}, 4} ={} \brk{k_2-q_3}^2, \quad
    &
    &D_{\texttt{dp2}, 5} ={} \brk{k_2+q_1+q_2}^2,\quad
    &
    &D_{\texttt{dp2}, 6} ={} \brk{k_1+k_2}^2,
    \nonumber\\*
    &D_{\texttt{dp2}, 7} ={} \brk{k_1+k_2+q_2}^2, \quad
    &
    &D_{\texttt{dp2}, 8} ={} k_2 \cdot q_2, \quad
    &
    &D_{\texttt{dp2}, 9} ={} k_1 \cdot q_3,
\\ \nonumber\\
    &D_{\texttt{dp3}, 1} ={} k_1^2,\quad
    &
    &D_{\texttt{dp3}, 2} ={} \brk{k_1-q_1}^2, \quad
    &
    &D_{\texttt{dp3}, 3} ={} \brk{k_1-q_1-q_2}^2,
    \nonumber\\*
    &D_{\texttt{dp3}, 4} ={} k_2^2, \quad
    &
    &D_{\texttt{dp3}, 5} ={} \brk{k_2-q_3}^2,\quad
    &
    &D_{\texttt{dp3}, 6} ={} \brk{k_1+k_2}^2,
    \nonumber\\*
    &D_{\texttt{dp3}, 7} ={} k_2\cdot q_1,\quad
    &
    &D_{\texttt{dp3}, 8} ={} k_2 \cdot q_2,\quad
    &
    &D_{\texttt{dp3}, 9} ={} k_1 \cdot q_3,
\\ \nonumber\\
    &D_{\texttt{dp4}, 1} ={} k_1^2,\quad
    &
    &D_{\texttt{dp4}, 2} ={} \brk{k_1-q_1}^2, \quad
    &
    &D_{\texttt{dp4}, 3} ={} \brk{k_1-q_1-q_2}^2,
    \nonumber\\*
    &D_{\texttt{dp4}, 4} ={} k_2^2, \quad
    &
    &D_{\texttt{dp4}, 5} ={} \brk{k_2+q_1+q_2+q_3}^2,\quad
    &
    &D_{\texttt{dp4}, 6} ={} \brk{k_1+k_2+q_3}^2,
    \nonumber\\*
    &D_{\texttt{dp4}, 7} ={} k_2\cdot q_2,\quad
    &
    &D_{\texttt{dp4}, 8} ={} k_1 \cdot q_3,\quad
    &
    &D_{\texttt{dp4}, 9} ={} k_2 \cdot q_3,
\\ \nonumber\\
    &D_{\texttt{dp5}, 1} ={} k_1^2,\quad
    &
    &D_{\texttt{dp5}, 2} ={} \brk{k_1-q_1}^2, \quad
    &
    &D_{\texttt{dp5}, 3} ={} k_2^2,
    \nonumber\\*
    &D_{\texttt{dp5}, 4} ={} \brk{k_2-q_3}^2, \quad
    &
    &D_{\texttt{dp5}, 5} ={} \brk{k_2+q_1+q_2}^2,\quad
    &
    &D_{\texttt{dp5}, 6} ={} \brk{k_1+k_2}^2,
    \nonumber\\*
    &D_{\texttt{dp5}, 7} ={} k_1\cdot q_2,\quad
    &
    &D_{\texttt{dp5}, 8} ={} k_2 \cdot q_2,\quad
    &
    &D_{\texttt{dp5}, 9} ={} k_1 \cdot q_3,
\\ \nonumber\\
    &D_{\texttt{dp6}, 1} ={} k_1^2,\quad
    &
    &D_{\texttt{dp6}, 2} ={} \brk{k_1-q_1}^2, \quad
    &
    &D_{\texttt{dp6}, 3} ={} k_2^2,
    \nonumber\\*
    &D_{\texttt{dp6}, 4} ={} \brk{k_2-q_3}^2, \quad
    &
    &D_{\texttt{dp6}, 5} ={} \brk{k_1+k_2}^2, \quad
    &
    &D_{\texttt{dp6}, 6} ={} \brk{k_1+k_2+q_2}^2,
    \nonumber\\*
    &D_{\texttt{dp6}, 7} ={} k_2\cdot q_1,\quad
    &
    &D_{\texttt{dp6}, 8} ={} k_2 \cdot q_2,\quad
    &
    &D_{\texttt{dp6}, 9} ={} k_1 \cdot q_3,
\\ \nonumber\\
    &D_{\texttt{dp7}, 1} ={} \brk{k_1-q_2}^2,\quad
    &
    &D_{\texttt{dp7}, 2} ={} \brk{k_1-q_1-q_2}^2, \quad
    &
    &D_{\texttt{dp7}, 3} ={} k_2^2,
    \nonumber\\*
    &D_{\texttt{dp7}, 4} ={} \brk{k_2-q_3}^2, \quad
    &
    &D_{\texttt{dp7}, 5} ={} \brk{k_2+q_1+q_2}^2,\quad
    &
    &D_{\texttt{dp7}, 6} ={} \brk{k_1+k_2}^2,
    \nonumber\\*
    &D_{\texttt{dp7}, 7} ={} k_1\cdot q_2,\quad
    &
    &D_{\texttt{dp7}, 8} ={} k_2 \cdot q_2,\quad
    &
    &D_{\texttt{dp7}, 9} ={} k_1 \cdot q_3,
\\ \nonumber\\
    &D_{\texttt{dp8}, 1} ={} \brk{k_1-q_3}^2,\quad
    &
    &D_{\texttt{dp8}, 2} ={} \brk{k_1-q_1-q_3}^2, \quad
    &
    &D_{\texttt{dp8}, 3} ={} k_2^2,
    \nonumber\\*
    &D_{\texttt{dp8}, 4} ={} \brk{k_2+q_1+q_2+q_3}^2, \quad
    &
    &D_{\texttt{dp8}, 5} ={} \brk{k_1+k_2}^2,\quad
    &
    &D_{\texttt{dp8}, 6} ={} \brk{k_1+k_2+q_2}^2,
    \nonumber\\*
    &D_{\texttt{dp8}, 7} ={} k_2\cdot q_2,\quad
    &
    &D_{\texttt{dp8}, 8} ={} k_1 \cdot q_3,\quad
    &
    &D_{\texttt{dp8}, 9} ={} k_2 \cdot q_3.
\end{alignat}

\subsection*{Pentabox of top-pair plus Higgs production}
\label{app:TIFamttH}
The TI families are pictorially represented in figure~\ref{fig:tifam-ttH}.
\begin{figure}[h]
    \centering
    \begin{subfigure}[b]{.3\textwidth}
        \centering
        \includegraphicsbox{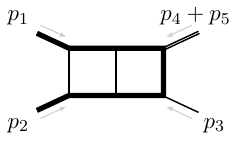}
        \caption{TI family \texttt{ttH7}}
    \end{subfigure}
    \begin{subfigure}[b]{.3\textwidth}
        \centering
        \includegraphicsbox{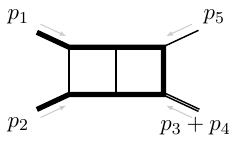}
        \caption{TI family \texttt{ttH8}}
    \end{subfigure}
    \\[.4cm]
    \foreach \i in {0,...,1} {%
        \begin{subfigure}[b]{.32\textwidth}
            \centering
            \includegraphicsbox{figures/ttH\the\numexpr3*\i+3\relax}
            \caption{TI family \texttt{ttH\the\numexpr3*\i+4\relax}}
        \end{subfigure}
        \begin{subfigure}[b]{.32\textwidth}
            \centering
            \includegraphicsbox{figures/ttH\the\numexpr3*\i+4\relax}
            \caption{TI family \texttt{ttH\the\numexpr3*\i+4\relax}}
        \end{subfigure}
        \begin{subfigure}[b]{.32\textwidth}
            \centering
            \includegraphicsbox{figures/ttH\the\numexpr3*\i+5\relax}
            \caption{TI family \texttt{ttH\the\numexpr3*\i+5\relax}}
        \end{subfigure}
        \\[.4cm]
    }
    \caption{\texttt{ttH} TI families.}
    \label{fig:tifam-ttH}
\end{figure}
The corresponding shifts of loop momenta read:
\begin{align}
    \texttt{ttH1},\>\texttt{ttH2},\texttt{ttH3},\texttt{ttH4}&:
    \quad
    \brk{k_1,\> k_2} \mapsto \brk{k_1,\> k_2},
     \nonumber\\
    \texttt{ttH5},\> &:
    \quad
    \brk{k_1,\> k_2} \mapsto \brk{k_1+p_3,\> k_2 - p_3},
     \nonumber\\
    \texttt{ttH6},\> \texttt{ttH7}&:
    \quad
    \brk{k_1,\> k_2} \mapsto \brk{k_1,\> k_2 - p_3},
    \nonumber\\
    \texttt{ttH8}&:
    \quad
    \brk{k_1,\> k_2} \mapsto \brk{k_1-p_2,\> k_2},
\end{align}
and the independent external momenta are:
\begin{align}
    \texttt{ttH1}&:
    \quad
    \brk{q_1,\> q_2,\> q_3} \defas \brk{p_1,\> p_2,\> p_3},
    \nonumber\\
    \texttt{ttH2}&:
    \quad
    \brk{q_1,\> q_2,\> q_3} \defas \brk{p_1,\> p_2,\> p_3+p_4},
     \nonumber\\
    \texttt{ttH3}&:
    \quad
    \brk{q_1,\> q_2,\> q_3} \defas \brk{p_1+p_2,\> p_3,\> p_4},
     \nonumber\\
    \texttt{ttH4}&:
    \quad
    \brk{q_1,\> q_2,\> q_3} \defas \brk{p_2,\> p_3,\> p_4},
     \nonumber\\
    \texttt{ttH5}&:
    \quad
    \brk{q_1,\> q_2,\> q_3} \defas \brk{p_1,\> p_2+p_3,\> p_4},
     \nonumber\\
    \texttt{ttH6}&:
    \quad
    \brk{q_1,\> q_2,\> q_3} \defas \brk{p_1,\> p_2,\> p_4},
     \nonumber\\
    \texttt{ttH7}&:
    \quad
    \brk{q_1,\> q_2,\> q_3} \defas \brk{p_2,\> p_1+p_3,\> p_4},
     \nonumber\\
    \texttt{ttH8}&:
    \quad
    \brk{q_1,\> q_2,\> q_3} \defas \brk{p_1,\> p_3,\> p_4}.
\end{align}
The generalized denominators of the TI families are:
{\allowdisplaybreaks
\begin{alignat}{3}
    &D_{\texttt{ttH1}, 1} ={} k_1^2-m_{\texttt{t}}^2,\quad
    &
    &D_{\texttt{ttH1}, 2} ={} \brk{k_1+q_2}^2,\,\,
    &
    &D_{\texttt{ttH1}, 3} ={} \brk{k_1+q_1+q_2}^2-m_{\texttt{t}}^2,
     \nonumber\\*
    &D_{\texttt{ttH1}, 4} ={} \brk{k_1+k_2}^2,\quad
     &
    &D_{\texttt{ttH1}, 5} ={} k_2^2-m_{\texttt{t}}^2,\,\,
    &
    &D_{\texttt{ttH1}, 6} ={} \brk{k_2+q_3}^2-m_{\texttt{t}}^2,
    \nonumber\\*
    &D_{\texttt{ttH1}, 7} ={} \brk{k_2-q_1-q_2}^2-m_{\texttt{t}}^2,\quad
    &
    &D_{\texttt{ttH1}, 8} ={} k_2\cdot q_2,\,\,
    &
    &D_{\texttt{ttH1}, 9} ={} k_1\cdot q_3,
    \\
    \nonumber
    \\
    &D_{\texttt{ttH2}, 1} ={} k_1^2-m_{\texttt{t}}^2,\quad
    &
    &D_{\texttt{ttH2}, 2} ={} \brk{k_1+q_2}^2,\,\,
    &
    &D_{\texttt{ttH2}, 3} ={} \brk{k_1+q_1+q_2}^2-m_{\texttt{t}}^2,
    \nonumber\\*
    &D_{\texttt{ttH2}, 4} ={} \brk{k_1+k_2}^2, \quad
    &
    &D_{\texttt{ttH2}, 5} ={} k_2^2-m_{\texttt{t}}^2,\,\,
    &
    &D_{\texttt{ttH2}, 6} ={} \brk{k_2+q_3}^2-m_{\texttt{t}}^2,
    \nonumber\\*
    &D_{\texttt{ttH2}, 7} ={} \brk{k_2-q_1-q_2}^2-m_{\texttt{t}}^2,\quad
    &
    &D_{\texttt{ttH2}, 8} ={} k_2\cdot q_2,\,\,
    &
    &D_{\texttt{ttH2}, 9} ={} k_1\cdot q_3,
    \\
    \nonumber
    \\
    &D_{\texttt{ttH3}, 1} ={} k_1^2-m_{\texttt{t}}^2,\quad
    &
    &D_{\texttt{ttH3}, 2} ={} \brk{k_1+q_2}^2-m_{\texttt{t}}^2,\,\,
    &
    &D_{\texttt{ttH3}, 3} ={} \brk{k_1+k_2}^2,
    \nonumber\\*
    &D_{\texttt{ttH3}, 4} ={} k_2^2-m_{\texttt{t}}^2,\quad
    &
    &D_{\texttt{ttH3}, 5} ={} \brk{k_2+q_2}^2-m_{\texttt{t}}^2,\,\,
    &
    &D_{\texttt{ttH3}, 6} ={} \brk{k_2+q_2+q_3}^2-m_{\texttt{t}}^2,
    \nonumber\\*
    &D_{\texttt{ttH3}, 7} ={} \brk{k_2-q_1}^2-m_{\texttt{t}}^2,\quad
    &
    &D_{\texttt{ttH3}, 8} ={} k_1\cdot q_2,\,\,
    &
    &D_{\texttt{ttH3}, 9} ={} k_1\cdot q_3,
    \\
    \nonumber
    \\
    &D_{\texttt{ttH4}, 1} ={} k_1^2-m_{\texttt{t}}^2,\quad
    &
    &D_{\texttt{ttH4}, 2} ={} \brk{k_1+q_1}^2,\,\,
    &
    &D_{\texttt{ttH4}, 3} ={} \brk{k_1+k_2}^2,
    \nonumber\\*
    &D_{\texttt{ttH4}, 4} ={} \brk{k_2}^2-m_{\texttt{t}}^2,\quad
    &
    &D_{\texttt{ttH4}, 5} ={} \brk{k_2+q_2}^2-m_{\texttt{t}}^2,\,\,
    &
    &D_{\texttt{ttH4}, 6} ={} \brk{k_2+q_2+q_3}^2-m_{\texttt{t}}^2,
    \nonumber\\*
    &D_{\texttt{ttH4}, 7} ={} k_2\cdot q_1,\quad
    &
    &D_{\texttt{ttH4}, 8} ={} k_1\cdot q_2,\,\,
    &
    &D_{\texttt{ttH4}, 9} ={} k_1\cdot q_3,
    \\
    \nonumber
    \\
    &D_{\texttt{ttH5}, 1} ={} \brk{k_1+q_2}^2,\quad
    &
    &D_{\texttt{ttH5}, 2} ={} \brk{k_1+q_1+q_2}^2-m_{\texttt{t}}^2,\,\,
    &
    &D_{\texttt{ttH5}, 3} ={} \brk{k_1+k_2}^2,
    \nonumber\\*
    &D_{\texttt{ttH5}, 4} ={} k_2^2-m_{\texttt{t}}^2,\quad
    &
    &D_{\texttt{ttH5}, 5} ={} \brk{k_2+q_3}^2-m_{\texttt{t}}^2,\,\,
    &
    &D_{\texttt{ttH5}, 6} ={} \brk{k_2-q_1-q_2}^2-m_{\texttt{t}}^2,
    \nonumber\\*
    &D_{\texttt{ttH5}, 7} ={} k_1\cdot q_2,\quad
    &
    &D_{\texttt{ttH5}, 8} ={} k_2\cdot q_2,\,\,
    &
    &D_{\texttt{ttH5}, 9} ={} k_1\cdot q_3,
    \\
    \nonumber
    \\
    &D_{\texttt{ttH6}, 1} ={} k_1^2-m_{\texttt{t}}^2,\quad
    &
    &D_{\texttt{ttH6}, 2} ={} \brk{k_1+q_2}^2,\,\,
    &
    &D_{\texttt{ttH6}, 3} ={} \brk{k_1+q_1+q_2}^2-m_{\texttt{t}}^2,
    \nonumber\\*
    &D_{\texttt{ttH6}, 4} ={} k_2^2-m_{\texttt{t}}^2,\quad
    &
    &D_{\texttt{ttH6}, 5} ={} \brk{k_2+q_3}^2-m_{\texttt{t}}^2,\,\,
    &
    &D_{\texttt{ttH6}, 6} ={} k_1\cdot k_2,
    \nonumber\\*
    &D_{\texttt{ttH6}, 7} ={} k_2\cdot q_1,\quad
    &
    &D_{\texttt{ttH6}, 8} ={} k_2\cdot q_2,\,\,
    &
    &D_{\texttt{ttH6}, 9} ={} k_1\cdot q_3,
    \\
    \nonumber
    \\
    &D_{\texttt{ttH7}, 1} ={} k_1^2-m_{\texttt{t}}^2,\quad
    &
    &D_{\texttt{ttH7}, 2} ={} \brk{k_1+q_2}^2,\,\,
    &
    &D_{\texttt{ttH7}, 3} ={} k_2^2-m_{\texttt{t}}^2,
    \nonumber\\*
    &D_{\texttt{ttH7}, 4} ={} \brk{k_2+q_3}^2-m_{\texttt{t}}^2, \quad
    &
    &D_{\texttt{ttH7}, 5} ={} \brk{k_2-q_1-q_2}^2-m_{\texttt{t}}^2,\,\,
    &
    &D_{\texttt{ttH7}, 6} ={} k_1\cdot k_2,
    \nonumber\\*
    &D_{\texttt{ttH7}, 7} ={} k_1\cdot q_2,\quad
    &
    &D_{\texttt{ttH7}, 8} ={} k_2\cdot q_2,\,\,
    &
    &D_{\texttt{ttH7}, 9} ={} k_1\cdot q_3,
    \\
    \nonumber
    \\
    &D_{\texttt{ttH8}, 1} ={} k_1^2,\quad
    &
    &D_{\texttt{ttH8}, 2} ={} \brk{k_1+q_1}^2-m_{\texttt{t}}^2,\,\,
    &
    &D_{\texttt{ttH8}, 3} ={} k_2^2-m_{\texttt{t}}^2,
    \nonumber\\*
    &D_{\texttt{ttH8}, 4} ={} \brk{k_2+q_2}^2-m_{\texttt{t}}^2, \quad
    &
    &D_{\texttt{ttH8}, 5} ={} \brk{k_2+q_2+q_3}^2-m_{\texttt{t}}^2,\,\,
    &
    &D_{\texttt{ttH8}, 6} ={} k_1\cdot k_2,
    \nonumber\\*
    &D_{\texttt{ttH8}, 7} ={} k_2\cdot q_1,\quad
    &
    &D_{\texttt{ttH8}, 8} ={} k_1\cdot q_2,\,\,
    &
    &D_{\texttt{ttH8}, 9} ={} k_1\cdot q_3,
\end{alignat}}
\subsection*{Massless two-loop six-point ladybug}
\label{app:TIFamLB}
\begin{figure}[H]
    \centering
    \foreach \i in {0,...,5} {%
        \begin{subfigure}[b]{.3\textwidth}
            \centering
            \includegraphicsbox{figures/lb\the\numexpr3*\i+1\relax}
            \caption{TI family \texttt{lb\the\numexpr3*\i+1\relax}}
            % \caption{\texttt{lb\the\numexpr3*\i+1\relax}}
        \end{subfigure}
        \begin{subfigure}[b]{.3\textwidth}
            \centering
            \includegraphicsbox{figures/lb\the\numexpr3*\i+2\relax}
            \caption{TI family \texttt{lb\the\numexpr3*\i+2\relax}}
            % \caption{\texttt{lb\the\numexpr3*\i+2\relax}}
        \end{subfigure}
        \begin{subfigure}[b]{.3\textwidth}
            \centering
            \includegraphicsbox{figures/lb\the\numexpr3*\i+3\relax}
            \caption{TI family \texttt{lb\the\numexpr3*\i+3\relax}}
            % \caption{\texttt{lb\the\numexpr3*\i+3\relax}}
        \end{subfigure}
        \\[.4cm]
    }
    \begin{subfigure}[b]{.3\textwidth}
        \centering
        \includegraphicsbox{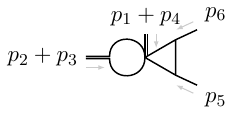}
        \caption{TI family \texttt{lb19}}
        % \caption{\texttt{lb19}}
    \end{subfigure}
    \caption{\texttt{lb} TI families}
    \label{fig:tifams-lb}
\end{figure}
The TI families are pictorially represented in figure~\ref{fig:tifams-lb}. The corresponding shifts of loop momenta read:
\begin{align}
    \texttt{lb1},\> \texttt{lb2}, \> \texttt{lb3}, \> \texttt{lb4}, \> \texttt{lb5}, \> \texttt{lb6}, \> \texttt{lb7}, \> \texttt{lb8}, \> \texttt{lb9}, \> \texttt{lb10}&:
    \quad
    \brk{k_1,\> k_2} \mapsto \brk{k_1,\> k_2},
     \nonumber\\
    \texttt{lb11},\> \texttt{lb12},\> \texttt{lb13},\> \texttt{lb18}&:
    \quad
    \brk{k_1,\> k_2} \mapsto \brk{k_1+p_1,\> k_2 - p_1},
     \nonumber\\
    \texttt{lb14}&:
    \quad
    \brk{k_1,\> k_2} \mapsto \brk{k_1 + p_1 + p_2,\> k_2 - p_1 - p_2},
     \nonumber\\
    \texttt{lb15},\> \texttt{lb16}&:
    \quad
    \brk{k_1,\> k_2} \mapsto \brk{k_1,\> k_2 - p_3},
      \nonumber\\
    \texttt{lb17},\> \texttt{lb19}&:
    \quad
    \brk{k_1,\> k_2} \mapsto \brk{k_1-p_4,\> k_2},
\end{align}
and the independent external momenta are:
\begin{align}
    \texttt{lb1}&:
    \quad
    \brk{q_1,\> q_2,\> q_3} \defas \brk{p_1,\> p_2,\> p_3},
    \nonumber\\
    \texttt{lb2}&:
    \quad
    \brk{q_1,\> q_2,\> q_3} \defas \brk{p_1,\> p_2+p_3,\> p_4+p_5},
     \nonumber\\
    \texttt{lb3}&:
    \quad
    \brk{q_1,\> q_2,\> q_3} \defas \brk{p_1,\> p_2+p_3,\> p_4},
     \nonumber\\
    \texttt{lb4}&:
    \quad
    \brk{q_1,\> q_2,\> q_3} \defas \brk{p_1+p_2,\> p_3,\> p_4+p_5},
     \nonumber\\
    \texttt{lb5}&:
    \quad
    \brk{q_1,\> q_2,\> q_3} \defas \brk{p_1+p_2,\> p_3,\> p_4},
     \nonumber\\
    \texttt{lb6}&:
    \quad
    \brk{q_1,\> q_2,\> q_3} \defas \brk{p_1+p_2+p_3,\> p_4,\> p_5},
     \nonumber\\
    \texttt{lb7}&:
    \quad
    \brk{q_1,\> q_2,\> q_3} \defas \brk{p_1,\> p_2,\> p_3+p_4+p_5},
     \nonumber\\
    \texttt{lb8}&:
    \quad
    \brk{q_1,\> q_2,\> q_3} \defas \brk{p_1,\> p_2,\> p_3+p_4},
      \nonumber\\
    \texttt{lb9}&:
    \quad
    \brk{q_1,\> q_2,\> q_3} \defas \brk{p_1,\> p_2+p_3+p_4,\> p_5},
     \nonumber\\
    \texttt{lb10}&:
    \quad
    \brk{q_1,\> q_2,\> q_3} \defas \brk{p_1+p_2,\> p_3+p_4,\> p_5},
     \nonumber\\
    \texttt{lb11}&:
    \quad
    \brk{q_1,\> q_2,\> q_3} \defas \brk{p_2,\> p_3,\> p_4+p_5},
     \nonumber\\
    \texttt{lb12}&:
    \quad
    \brk{q_1,\> q_2,\> q_3} \defas \brk{p_2,\> p_3,\> p_4},
     \nonumber\\
    \texttt{lb13}&:
    \quad
    \brk{q_1,\> q_2,\> q_3} \defas \brk{p_2+p_3,\> p_4,\> p_5},
     \nonumber\\
    \texttt{lb14}&:
    \quad
    \brk{q_1,\> q_2,\> q_3} \defas \brk{p_3,\> p_4,\> p_5},
     \nonumber\\
    \texttt{lb15}&:
    \quad
    \brk{q_1,\> q_2,\> q_3} \defas \brk{p_1,\> p_2,\> p_4+p_5},
     \nonumber\\
    \texttt{lb16}&:
    \quad
    \brk{q_1,\> q_2,\> q_3} \defas \brk{p_1+p_2,\> p_4,\> p_5},
     \nonumber\\
    \texttt{lb17}&:
    \quad
    \brk{q_1,\> q_2,\> q_3} \defas \brk{p_2,\> p_3,\> p_1+p_4},
     \nonumber\\
    \texttt{lb18}&:
    \quad
   \brk{q_1,\> q_2,\> q_3} \defas \brk{p_2,\> p_3+p_4,\> p_5},
    \nonumber\\
    \texttt{lb19}&:
    \quad
    \brk{q_1,\> q_2,\> q_3} \defas \brk{p_2+p_3,\> p_1+p_4,\> p_5}.
\end{align}
The generalized denominators of the TI families are:
{\allowdisplaybreaks
\begin{alignat}{3}
    &D_{\texttt{lb1}, 1} ={} k_1^2,\quad
    &
    &D_{\texttt{lb1}, 2} ={} \brk{k_1-q_1}^2,\quad
    &
    &D_{\texttt{lb1}, 3} ={} \brk{k_1-q_1-q_2}^2,
    \nonumber\\*
    &D_{\texttt{lb1}, 4} ={} \brk{k_1-q_1-q_2-q_3}^2,\quad
    &
    &D_{\texttt{lb1}, 5} ={} k_2^2,\quad
    &
    &D_{\texttt{lb1}, 6} ={}\brk{k_2+q_1+q_2+q_3}^2,\nonumber\\*
    &D_{\texttt{lb1}, 7} ={} \brk{k_1+k_2}^2, \quad
    &
    &D_{\texttt{lb1}, 8} ={} k_2\cdot q_2, \quad
    &
    &D_{\texttt{lb1}, 9} ={} k_2\cdot q_3, \quad
\\ \nonumber\\
    &D_{\texttt{lb2}, 1} ={} k_1^2,\quad
    &
    &D_{\texttt{lb2}, 2} ={} \brk{k_1-q_1}^2,\quad
    &
    &D_{\texttt{lb2}, 3} ={} \brk{k_1-q_1-q_2}^2,
    \nonumber\\*
    &D_{\texttt{lb2}, 4} ={} k_2^2,\quad
    &
    &D_{\texttt{lb2}, 5} ={} \brk{k_2+q_1+q_2+q_3}^2,\quad
    &
    &D_{\texttt{lb2}, 6} ={}\brk{k_2+q_1+q_2}^2,\nonumber\\*
    &D_{\texttt{lb2}, 7} ={} \brk{k_1+k_2}^2,\quad
    &
    &D_{\texttt{lb2}, 8} ={} k_2\cdot q_2, \quad
    &
    &D_{\texttt{lb2}, 9} ={} k_1\cdot q_3, \quad
\\ \nonumber\\
    &D_{\texttt{lb3}, 1} ={} k_1^2,\quad
    &
    &D_{\texttt{lb3}, 2} ={} \brk{k_1-q_1}^2,\quad
    &
    &D_{\texttt{lb3}, 3} ={} \brk{k_1-q_1-q_2}^2,
    \nonumber\\*
    &D_{\texttt{lb3}, 4} ={} k_2^2,\quad
    &
    &D_{\texttt{lb3}, 5} ={} \brk{k_2+q_1+q_2+q_3}^2,\quad
    &
    &D_{\texttt{lb3}, 6} ={}\brk{k_2+q_1+q_2}^2,\nonumber\\*
    &D_{\texttt{lb3}, 7} ={} \brk{k_1+k_2}^2, \quad
    &
    &D_{\texttt{lb3}, 8} ={} k_2\cdot q_2, \quad
    &
    &D_{\texttt{lb3}, 9} ={} k_1\cdot q_3, \quad
\\ \nonumber\\
    &D_{\texttt{lb4}, 1} ={} k_1^2,\quad
    &
    &D_{\texttt{lb4}, 2} ={} \brk{k_1-q_1}^2,\quad
    &
    &D_{\texttt{lb4}, 3} ={} \brk{k_1-q_1-q_2}^2,
    \nonumber\\*
    &D_{\texttt{lb4}, 4} ={} k_2^2,\quad
    &
    &D_{\texttt{lb4}, 5} ={} \brk{k_2+q_1+q_2+q_3}^2,\quad
    &
    &D_{\texttt{lb4}, 6} ={}\brk{k_2+q_1+q_2}^2,\nonumber\\*
    &D_{\texttt{lb4}, 7} ={} \brk{k_1+k_2}^2, \quad
    &
    &D_{\texttt{lb4}, 8} ={} k_2\cdot q_2, \quad
    &
    &D_{\texttt{lb4}, 9} ={} k_1\cdot q_3, \quad
\\ \nonumber\\
    &D_{\texttt{lb5}, 1} ={} k_1^2,\quad
    &
    &D_{\texttt{lb5}, 2} ={} \brk{k_1-q_1}^2,\quad
    &
    &D_{\texttt{lb5}, 3} ={} \brk{k_1-q_1-q_2}^2,
    \nonumber\\*
    &D_{\texttt{lb5}, 4} ={} k_2^2,\quad
    &
    &D_{\texttt{lb5}, 5} ={} \brk{k_2+q_1+q_2+q_3}^2,\quad
    &
    &D_{\texttt{lb5}, 6} ={}\brk{k_2+q_1+q_2}^2,\nonumber\\*
    &D_{\texttt{lb5}, 7} ={} \brk{k_1+k_2}^2, \quad
    &
    &D_{\texttt{lb5}, 8} ={} k_2\cdot q_2, \quad
    &
    &D_{\texttt{lb5}, 9} ={} k_1\cdot q_3, \quad
\\ \nonumber\\
    &D_{\texttt{lb6}, 1} ={} k_1^2,\quad
    &
    &D_{\texttt{lb6}, 2} ={} \brk{k_1-q_1}^2,\quad
    &
    &D_{\texttt{lb6}, 3} ={} k_2^2,
    \nonumber\\*
    &D_{\texttt{lb6}, 4} ={} \brk{k_2+q_1+q_2+q_3}^2,\quad
    &
    &D_{\texttt{lb6}, 5} ={} \brk{k_2+q_1+q_2}^2,\quad
    &
    &D_{\texttt{lb6}, 6} ={}\brk{k_2+q_1}^2,\nonumber\\*
    &D_{\texttt{lb6}, 7} ={} \brk{k_1+k_2}^2, \quad
    &
    &D_{\texttt{lb6}, 8} ={} k_1\cdot q_2, \quad
    &
    &D_{\texttt{lb6}, 9} ={} k_1\cdot q_3, \quad
\\ \nonumber\\
    &D_{\texttt{lb7}, 1} ={} k_1^2,\quad
    &
    &D_{\texttt{lb7}, 2} ={} \brk{k_1-q_1}^2,\quad
    &
    &D_{\texttt{lb7}, 3} ={} \brk{k_1-q_1-q_2}^2,
    \nonumber\\*
    &D_{\texttt{lb7}, 4} ={} k_2^2,\quad
    &
    &D_{\texttt{lb7}, 5} ={} \brk{k_2+q_1+q_2+q_3}^2,\quad
    &
    &D_{\texttt{lb7}, 6} ={}\brk{k_1+k_2}^2,\nonumber\\*
    &D_{\texttt{lb7}, 7} ={} k_2\cdot q_2, \quad
    &
    &D_{\texttt{lb7}, 8} ={} k_1\cdot q_3, \quad
    &
    &D_{\texttt{lb7}, 9} ={} k_2\cdot q_3, \quad
\\ \nonumber\\
    &D_{\texttt{lb8}, 1} ={} k_1^2,\quad
    &
    &D_{\texttt{lb8}, 2} ={} \brk{k_1-q_1}^2,\quad
    &
    &D_{\texttt{lb8}, 3} ={} \brk{k_1-q_1-q_2}^2,
    \nonumber\\*
    &D_{\texttt{lb8}, 4} ={} k_2^2,\quad
    &
    &D_{\texttt{lb8}, 5} ={} \brk{k_2+q_1+q_2+q_3}^2,\quad
    &
    &D_{\texttt{lb8}, 6} ={}\brk{k_1+k_2}^2,\nonumber\\*
    &D_{\texttt{lb8}, 7} ={} k_2\cdot q_2, \quad
    &
    &D_{\texttt{lb8}, 8} ={} k_1\cdot q_3, \quad
    &
    &D_{\texttt{lb8}, 9} ={} k_2\cdot q_3, \quad
\\ \nonumber\\
    &D_{\texttt{lb9}, 1} ={} k_1^2,\quad
    &
    &D_{\texttt{lb9}, 2} ={} \brk{k_1-q_1}^2,\quad
    &
    &D_{\texttt{lb9}, 3} ={} k_2^2,
    \nonumber\\*
    &D_{\texttt{lb9}, 4} ={} \brk{k_2+q_1+q_2+q_3}^2,\quad
    &
    &D_{\texttt{lb9}, 5} ={} \brk{k_2+q_1+q_2}^2,\quad
    &
    &D_{\texttt{lb9}, 6} ={}\brk{k_1+k_2}^2,\nonumber\\*
    &D_{\texttt{lb9}, 7} ={} k_1\cdot q_2, \quad
    &
    &D_{\texttt{lb9}, 8} ={} k_2\cdot q_2, \quad
    &
    &D_{\texttt{lb9}, 9} ={} k_1\cdot q_3, \quad
\\ \nonumber\\
    &D_{\texttt{lb10}, 1} ={} k_1^2,\quad
    &
    &D_{\texttt{lb10}, 2} ={} \brk{k_1-q_1}^2,\quad
    &
    &D_{\texttt{lb10}, 3} ={} k_2^2,
    \nonumber\\*
    &D_{\texttt{lb10}, 4} ={} \brk{k_2+q_1+q_2+q_3}^2,\quad
    &
    &D_{\texttt{lb10}, 5} ={} \brk{k_2+q_1+q_2}^2,\quad
    &
    &D_{\texttt{lb10}, 6} ={}\brk{k_1+k_2}^2,\nonumber\\*
    &D_{\texttt{lb10}, 7} ={} k_1\cdot q_2, \quad
    &
    &D_{\texttt{lb10}, 8} ={} k_2\cdot q_2, \quad
    &
    &D_{\texttt{lb10}, 9} ={} k_1\cdot q_3, \quad
\\ \nonumber\\
    &D_{\texttt{lb11}, 1} ={} k_1^2,\quad
    &
    &D_{\texttt{lb11}, 2} ={} \brk{k_1-q_1}^2,\quad
    &
    &D_{\texttt{lb11}, 3} ={} \brk{k_1-q_1-q_2}^2,
    \nonumber\\*
    &D_{\texttt{lb11}, 4} ={} \brk{k_2+q_1+q_2+q_3}^2,\quad
    &
    &D_{\texttt{lb11}, 5} ={} \brk{k_2+q_1+q_2}^2,\quad
    &
    &D_{\texttt{lb11}, 6} ={}\brk{k_1+k_2}^2,\nonumber\\*
    &D_{\texttt{lb11}, 7} ={} k_2\cdot q_1, \quad
    &
    &D_{\texttt{lb11}, 8} ={} k_2\cdot q_2, \quad
    &
    &D_{\texttt{lb11}, 9} ={} k_1\cdot q_3, \quad
\\ \nonumber\\
    &D_{\texttt{lb12}, 1} ={} k_1^2,\quad
    &
    &D_{\texttt{lb12}, 2} ={} \brk{k_1-q_1}^2,\quad
    &
    &D_{\texttt{lb12}, 3} ={} \brk{k_1-q_1-q_2}^2,
    \nonumber\\*
    &D_{\texttt{lb12}, 4} ={} \brk{k_2+q_1+q_2+q_3}^2,\quad
    &
    &D_{\texttt{lb12}, 5} ={} \brk{k_2+q_1+q_2}^2,\quad
    &
    &D_{\texttt{lb12}, 6} ={}\brk{k_1+k_2}^2,\nonumber\\*
    &D_{\texttt{lb12}, 7} ={} k_2\cdot q_1, \quad
    &
    &D_{\texttt{lb12}, 8} ={} k_2\cdot q_2, \quad
    &
    &D_{\texttt{lb12}, 9} ={} k_1\cdot q_3, \quad
\\ \nonumber\\
    &D_{\texttt{lb13}, 1} ={} k_1^2,\quad
    &
    &D_{\texttt{lb13}, 2} ={} \brk{k_1-q_1}^2,\quad
    &
    &D_{\texttt{lb13}, 3} ={} \brk{k_2+q_1+q_2+q_3}^2,
    \nonumber\\*
    &D_{\texttt{lb13}, 4} ={} \brk{k_2+q_1+q_2}^2,\quad
    &
    &D_{\texttt{lb13}, 5} ={} \brk{k_2+q_1}^2,\quad
    &
    &D_{\texttt{lb13}, 6} ={}\brk{k_1+k_2}^2,\nonumber\\*
    &D_{\texttt{lb13}, 7} ={} k_2\cdot q_1, \quad
    &
    &D_{\texttt{lb13}, 8} ={} k_1\cdot q_2, \quad
    &
    &D_{\texttt{lb13}, 9} ={} k_1\cdot q_3, \quad
\\ \nonumber\\
    &D_{\texttt{lb14}, 1} ={} k_1^2,\quad
    &
    &D_{\texttt{lb14}, 2} ={} \brk{k_1-q_1}^2,\quad
    &
    &D_{\texttt{lb14}, 3} ={} \brk{k_2+q_1+q_2+q_3}^2,
    \nonumber\\*
    &D_{\texttt{lb14}, 4} ={} \brk{k_2+q_1+q_2}^2,\quad
    &
    &D_{\texttt{lb14}, 5} ={} \brk{k_2+q_1}^2,\quad
    &
    &D_{\texttt{lb14}, 6} ={}\brk{k_1+k_2}^2,\nonumber\\*
    &D_{\texttt{lb14}, 7} ={} k_2\cdot q_1, \quad
    &
    &D_{\texttt{lb14}, 8} ={} k_1\cdot q_2, \quad
    &
    &D_{\texttt{lb14}, 9} ={} k_1\cdot q_3, \quad
\\ \nonumber\\
    &D_{\texttt{lb15}, 1} ={} k_1^2,\quad
    &
    &D_{\texttt{lb15}, 2} ={} \brk{k_1-q_1}^2,\quad
    &
    &D_{\texttt{lb15}, 3} ={} \brk{k_1-q_1-q_2}^2,
    \nonumber\\*
    &D_{\texttt{lb15}, 4} ={} \brk{k_2+q_1+q_2+q_3}^2,\quad
    &
    &D_{\texttt{lb15}, 5} ={} \brk{k_2+q_1+q_2}^2,\quad
    &
    &D_{\texttt{lb15}, 6} ={} k_1\cdot k_2,\nonumber\\*
    &D_{\texttt{lb15}, 7} ={} k_2\cdot q_1, \quad
    &
    &D_{\texttt{lb15}, 8} ={} k_2\cdot q_2, \quad
    &
    &D_{\texttt{lb15}, 9} ={} k_1\cdot q_3, \quad
\\ \nonumber\\
    &D_{\texttt{lb16}, 1} ={} k_1^2,\quad
    &
    &D_{\texttt{lb16}, 2} ={} \brk{k_1-q_1}^2,\quad
    &
    &D_{\texttt{lb16}, 3} ={} \brk{k_2+q_1+q_2+q_3}^2,
    \nonumber\\*
    &D_{\texttt{lb16}, 4} ={} \brk{k_2+q_1+q_2}^2,\quad
    &
    &D_{\texttt{lb16}, 5} ={} \brk{k_2+q_1}^2,\quad
    &
    &D_{\texttt{lb16}, 6} ={} k_1\cdot k_2,\nonumber\\*
    &D_{\texttt{lb16}, 7} ={} k_2\cdot q_1, \quad
    &
    &D_{\texttt{lb16}, 8} ={} k_1\cdot q_2, \quad
    &
    &D_{\texttt{lb16}, 9} ={} k_1\cdot q_3, \quad
\\ \nonumber\\
    &D_{\texttt{lb17}, 1} ={} \brk{k_1-q_3}^2,\quad
    &
    &D_{\texttt{lb17}, 2} ={} \brk{k_1-q_1-q_3}^2,\quad
    &
    &D_{\texttt{lb17}, 3} ={} \brk{k_1-q_1-q_2-q_3}^2,
    \nonumber\\*
    &D_{\texttt{lb17}, 4} ={} k_2^2,\quad
    &
    &D_{\texttt{lb17}, 5} ={} \brk{k_2+q_1+q_2+q_3}^2,\quad
    &
    &D_{\texttt{lb17}, 6} ={} k_1\cdot k_2,\nonumber\\*
    &D_{\texttt{lb17}, 7} ={} k_2\cdot q_2, \quad
    &
    &D_{\texttt{lb17}, 8} ={} k_1\cdot q_3, \quad
    &
    &D_{\texttt{lb17}, 9} ={} k_2\cdot q_3, \quad
\\ \nonumber\\
    &D_{\texttt{lb18}, 1} ={} k_1^2,\quad
    &
    &D_{\texttt{lb18}, 2} ={} \brk{k_1-q_1}^2,\quad
    &
    &D_{\texttt{lb18}, 3} ={} \brk{k_2+q_1+q_2+q_3}^2,
    \nonumber\\*
    &D_{\texttt{lb18}, 4} ={} \brk{k_2+q_1+q_2}^2,\quad
    &
    &D_{\texttt{lb18}, 5} ={} \brk{k_1+k_2}^2,\quad
    &
    &D_{\texttt{lb18}, 6} ={} k_2\cdot q_1,\nonumber\\*
    &D_{\texttt{lb18}, 7} ={} k_1\cdot q_2, \quad
    &
    &D_{\texttt{lb18}, 8} ={} k_2\cdot q_2, \quad
    &
    &D_{\texttt{lb18}, 9} ={} k_2\cdot q_3, \quad
\\ \nonumber\\
    &D_{\texttt{lb19}, 1} ={} \brk{k_1-q_2}^2,\quad
    &
    &D_{\texttt{lb19}, 2} ={} \brk{k_1-q_1-q_2}^2,\quad
    &
    &D_{\texttt{lb19}, 3} ={} k_2^2,
    \nonumber\\*
    &D_{\texttt{lb19}, 4} ={} \brk{k_2+q_1+q_2+q_3}^2,\quad
    &
    &D_{\texttt{lb19}, 5} ={} \brk{k_2+q_1+q_2}^2,\quad
    &
    &D_{\texttt{lb19}, 6} ={} k_1\cdot k_2,\nonumber\\*
    &D_{\texttt{lb19}, 7} ={} k_1\cdot q_2, \quad
    &
    &D_{\texttt{lb19}, 8} ={} k_2\cdot q_2, \quad
    &
    &D_{\texttt{lb19}, 9} ={} k_2\cdot q_3. \quad
\end{alignat}}}
\bibliographystyle{JHEP}
\bibliography{biblio}
\end{document}